\documentclass[11pt]{article}
\usepackage[english]{babel}
\usepackage[T1]{fontenc}
\usepackage{amsmath,amsthm,amsfonts,amssymb,amsopn,amscd, mathrsfs}
\usepackage{cite}

\newlength{\dinwidth}
\newlength{\dinmargin}
\newcommand{\pir}{\pi_{\r}}
\newcommand{\pil}{\pi_{\l}} 
\newcommand{\V}{U^{\mathrm{char}} }
\newcommand{\mcW}{\mathcal W}
\newcommand{\bq}{\boldsymbol q}
\newcommand{\M}{\mathcal M}
\newcommand{\hh}{\mathfrak{H}}
\newcommand{\kk}{\mathfrak{K}}
\newcommand{\bQ}{\boldsymbol Q}
\newcommand{\II}{\mathbb{I}}
\newcommand{\JJ}{\mathbb{J}}
\newcommand{\rr}{\mathrm{R}}
\renewcommand{\ll}{\mathrm{L}}

\newcommand{\FF}{G}
\newcommand{\Ltt}{{\hat{L}}}
\newcommand{\Att}{\hat{A}}
\newcommand{\mcLtt}{\hat{\mcL}}
\newcommand{\hiltt}{\hat{\hil}}
\newcommand{\Btt}{\hat\B}
\newcommand{\Ut}{\hat\U}
\newcommand{\At}{\hat\A}
\newcommand{\BttvN}{\Btt^{\textrm{vN} }}
\newcommand{\Utt}{\hat U}
\newcommand{\hA}{\bar\A}

\newcommand{\Sub}{C}
\newcommand{\Dom}{\mathcal{D}}
\newcommand{\Omz}{\Om_0}
\renewcommand{\t}{s}
\newcommand{\pb}{\boldsymbol p}   
\newcommand{\xb}{\boldsymbol x}   
\newcommand{\Pb}{\boldsymbol P }
\renewcommand{\r}{{\rm{R}}}
\renewcommand{\l}{{\rm{L}}}
\newcommand{\BvN}{\B^{\textrm{vN} } }
\newcommand{\hilk}{\mathcal{K}} 
\newcommand{\alb}{\be}
\newcommand{\U}{V}

\newcommand{\I}{\mathfrak{I}}
\newcommand{\J}{\mathfrak{J}}
\newcommand{\can}{\mathrm{can}}

\newcommand{\odd}{\mathrm{odd}}
\newcommand{\ev}{\mathrm{ev}}

\newcommand{\tf}{\tilde f}

\newcommand{\De}{\Delta}
\newcommand{\mcL}{\mathcal{L}}
\newcommand{\mcC}{\mathcal{C}}
\newcommand{\ti}{\tilde}
\newcommand{\slim}{\te{s}\textrm{-}\lim}
\newcommand{\pout}{\overset{\tout}{\times}}
\newcommand{\pin}{\overset{\tin}{\times}}
\newcommand{\tin}{\te{in}}
\newcommand{\tout}{\te{ out}}
\newcommand{\Phip}{\Phi^{\tout}_+}
\newcommand{\Phim}{\Phi^{\tout}_-}
\newcommand{\ph}{\phantom}
\newcommand{\x}{x}
\newcommand{\ran}{\rangle}
\newcommand{\lan}{\langle}

\newcommand{\wt}{\widetilde}
\newcommand{\Om}{\Omega}
\newcommand{\si}{\sigma}

\newcommand{\h}{\fr{1}{2}}

\newcommand{\te}{\mathrm}

\newcommand{\nat}{\mathbb{N}}
\newcommand{\hil}{\mathcal{H}}
\newcommand{\om}{\omega}
\newcommand{\mfa}{\mathfrak{A}}

\newcommand{\mco}{\mathcal{O}}

\newcommand{\eps}{\varepsilon}
\newcommand{\fr}[2]{\frac{#1}{#2}}
\newcommand{\al}{\alpha}
\newcommand{\be}{\beta}

\newcommand{\real}{\mathbb{R}}

\newcommand{\ov}{\overline}

\newcommand{\ga}{\gamma}
\newcommand{\non}{\nonumber}

\def\proof{\noindent{\bf Proof. }}
\def\qed{$\Box$\medskip}

\newtheorem{theoreme}{Theorem } [section]
\newtheorem{proposition}[theoreme]{Proposition}
\newtheorem{lemma}[theoreme]{Lemma}
\newtheorem{definition}[theoreme]{Definition}
\newtheorem{corollary}[theoreme]{Corollary}
\newtheorem{example}[theoreme]{Example}
\newtheorem{criterion}[theoreme]{Criterion}
\theoremstyle{remark}
\newtheorem{remark}[theoreme]{Remark} 
\newcommand{\beq}{\begin{equation}}
\newcommand{\eeq}{\end{equation}}
\newcommand{\beqa}{\begin{eqnarray}}
\newcommand{\eeqa}{\end{eqnarray}}
\newcommand{\ben}{\begin{arabicenumerate}}
\newcommand{\een}{\end{arabicenumerate}}
\newcommand{\bex}{\begin{example}}
\newcommand{\eex}{\end{example}}
\newcommand{\ber}{\begin{remark}}
\newcommand{\eer}{\end{remark}}
\newcommand{\bec}{\begin{corollary}}
\newcommand{\eec}{\end{corollary}}
\newcommand{\bed}{\begin{definition}}
\newcommand{\eed}{\end{definition}}
\newcommand{\bep}{\begin{proposition}}
\newcommand{\eep}{\end{proposition}}
\newcommand{\becr}{\begin{criterion}}
\newcommand{\eecr}{\end{criterion}}
\def\bel{\begin{lemma}}
\def\eel{\end{lemma}}
\def\bet{\begin{theoreme}}
\def\eet{\end{theoreme}}
\def\bed{\begin{definition}}
\def\eed{\end{definition}}

\def\RR{{\mathbb R}}
\def\CC{{\mathbb C}}

\def\ZZ{{\mathbb Z}}

\def\ti{\tilde}
\def\lan{\langle}
\def\ran{\rangle}
\def\Ad{{\hbox{\rm Ad}}}

\def\id{{\rm id}}

\def\si{\sigma}

\def\om{\omega}
\def\Om{\Omega}

\def\A{{\cal A}}
\def\B{{\mfa}}

\def\M{{\cal M}}

\def\O{{\cal O}}

\def\I{{\cal I}}
\def\H{{\cal H}}
\def\K{{\cal K}}
\def\S{{\cal S}}

\def\emptyset{\varnothing}

\def\S2{S^{1(2)}}

\def\<{\langle}
\def\>{\rangle}

\def\supp{\mathrm{supp}}

\begin{document}

\title{Infraparticles with superselected direction of motion in two-dimensional conformal field theory}

\author{
{\bf Wojciech Dybalski\footnote{Supported by the DFG grant SP181/25.}}\\
Zentrum Mathematik, Technische Universit\"at M\"unchen,\\
D-85747 Garching, Germany\\
E-mail: {\tt dybalski@ma.tum.de}
\and
{\bf Yoh Tanimoto\footnote{Supported in part by the ERC Advanced Grant 227458
OACFT ``Operator Algebras and Conformal Field Theory''.}}\\
Dipartimento di Matematica, Universit\`a di Roma ``Tor
Vergata''\\ Via della Ricerca Scientifica, 1 - I--00133 Roma, Italy.\\
E-mail: {\tt tanimoto@mat.uniroma2.it}}

\date{ }
\maketitle

\begin{abstract}

Particle aspects of two-dimensional conformal field theories are investigated,
using methods from algebraic quantum field theory. 
The results include asymptotic completeness in terms of (counterparts of) Wigner particles in 
any vacuum representation and the existence of (counterparts of)
infraparticles in any charged irreducible product representation
of a given chiral conformal field theory. Moreover, an interesting interplay between
the infraparticle's direction of motion and the superselection structure is demonstrated 
in a large class of examples. This phenomenon resembles the
electron's momentum superselection 
expected in quantum electrodynamics.

\end{abstract}

\section{Introduction}\label{intro}
\setcounter{equation}{0}

Particle aspects and superselection structure of quantum electrodynamics
are plagued by the infrared problem, which has been a subject of
study in mathematical physics for more than four decades \cite{Sch63,Fr73,Fr74.1,Pi03,DG04,Pi05,CFP07,HH08,CFP09,FMS79, Bu77, Bu82, Bu86, Bu90, BPS91, Po04.1, Po04.2,Dy05,Dy08.3,Dy10, St00, He07,Re09 }. 
The origin of this difficulty, inherited from classical electrodynamics, is the emission
of  photons which accompanies any change of the electron's momentum. It has two
important consequences which are closely related: Firstly, the electron is not a particle in the sense of Wigner \cite{Wi39}, 
but rather an \emph{infraparticle} \cite{Sch63} i.e. it does 
not have a precise mass. Secondly, the electron's plane wave configurations of different momenta cannot be
superposed into normalizable wavepackets. In fact, such configurations have different spacelike  asymptotic flux of the
electric field, which imposes a superselection rule \cite{Bu82}. 
The evidence for this  phenomenon of the electron's
\emph{momentum superselection} comes from two sources:
On the one hand, it appears in  models of non-relativistic
QED  in the representation structure of the asymptotic electromagnetic field algebra \cite{CFP07}. 
On the other hand, it is suggested by structural 
results in the general framework of algebraic quantum field theory \cite{Bu82,Bu86,BPS91,Po04.1,Po04.2}. However, no examples of
local, relativistic theories, describing infraparticles with superselected momentum, have been given to date. 
Thus the logical consistency of this property with the basic postulates of quantum field theory
remains to be settled. As a step in this direction, we demonstrate in the present paper that a simple variant of this 
phenomenon - superselection of  direction of motion - occurs in a large class of two-dimensional conformal field theories.

Conformal field theory has been a subject of intensive research over the last two
decades, both from physical and mathematical viewpoints, motivated, in particular, 
by the search for non-trivial quantum field theories. (See e.g. \cite{BMRW09} and references therein). 
It exhibits
particularly interesting properties in two dimensions, where the symmetry group 
is infinite dimensional. Since the seminal work
of Buchholz, Mack and Todorov \cite{BMT88} the superselection structure of these
theories has been investigated \cite{GF93} and deep classification results have been obtained \cite{ KL04.1,KL04.2}.
It has remained unnoticed, however, that  two-dimensional conformal field 
theories have also a rich and interesting particle structure:
The concepts of Wigner particles 
and infraparticles have their natural counterparts in this setting and both types
of excitations appear in abundance: Any chiral conformal field theory in a vacuum
representation has a complete particle interpretation in terms of Wigner particles.
Although such theories are non-interacting, their (Grosse-Lechner) deformations \cite{BLS10} 
exhibit non-trivial scattering and inherit the property of asymptotic completeness as we show in a companion paper~\cite{DT10}.
It is verified in the present work that any charged irreducible product representation of a chiral conformal field theory admits infraparticles. In a large class 
of examples these infraparticles have superselected direction of motion i.e. their plane wave configurations
with opposite directions of momentum cannot be superposed. Thus subtle particle phenomena, which are not 
under control in physical spacetime, can be investigated in these two-dimensional models.

To keep our analysis general, we rely on the setting of algebraic QFT \cite{Ha}. We base
our discussion on the concept of a local net of   $C^*$-algebras on $\real^2$, defined
precisely in Subsection~\ref{first-preliminaries}: To
any open, bounded region $\mco\subset\real^2$ we attach a   $C^*$-algebra $\B(\mco)$,
acting on a Hilbert space $\hil$ of physical states. This algebra is  generated by observables which can be measured with an experimental device localized in $\mco$.
It is contained in the quasilocal algebra $\B$, which is the inductive limit of the net $\mco\to\B(\mco)$.  
Moreover, there acts a unitary representation of translations $\real^2\ni x\to U(x)$ on $\hil$,
whose adjoint action $\al_{x}(\,\cdot\,)=U(x)\,\cdot\,U(x)^*$ shifts the observables in spacetime. 
The infinitesimal generators of $U$ are interpreted as the Hamiltonian $H$ and the momentum
operator $\Pb$. Their joint spectrum is contained in the closed forward lightcone $V_+$, to ensure the positivity
of energy. If there exists  a cyclic, unit vector $\Om\in\hil$ 
which is the unique (up to a phase)  joint eigenvector of  $H$ and $\Pb$ with eigenvalue zero, then we say that the theory
is in a vacuum representation. If each of the subspaces
$\hil_{\pm}=\ker(H\mp \Pb)$ includes some vectors orthogonal to $\Om$, then we say that the theory contains
Wigner particles. Since we do not
assume that these particles are described by vectors in some irreducible representation space of the
Poincar\'e group,  the present definition is less restrictive than the conventional one. However, it is better suited for a
description of the dispersionless kinematics of two-dimensional massless excitations. In particular, it allows
us to apply the natural scattering theory, developed by Buchholz in \cite{Bu75}, which we outline in Subsection~\ref{Wigner} below.
We recall that in \cite{Bu75} these excitations are called `waves', to stress their composite character.

Due to this compound structure of Wigner particles (or `waves'), asymptotic completeness in a
vacuum representation is not in conflict with the existence of charged representations with a non-trivial particle content.
However, in charged representations of massless two-dimensional theories Wigner particles may be absent, as noticed in \cite{Bu96}.
In this case  scattering theory from \cite{Bu75} does not apply and an appropriate framework for the analysis of  particle aspects 
is the theory of \emph{particle weights} \cite{Bu90,St89, BPS91,Jo91, Po04.1,Po04.2,Dy08.3,Dy10}, developed by Buchholz, Porrmann and Stein, which we revisit in Subsections~\ref{weights} and \ref{asymptotic-functionals-sub}. This theory is based on the concept of the asymptotic functional, given by
\beq
\si_{\Psi}^{\tout}(C)=\lim_{t\to\infty}\int d\xb\, (\Psi|\al_{(t,\xb)}(C)\Psi), \label{asymptotic-functional-first}
\eeq
for any vector $\Psi\in\hil$ of bounded energy, and suitable observables $C\in\B$. (In general,
some time averaging  and restriction to a subnet may be needed before taking the limit).
We remark for future reference that this functional
induces a sesquilinear form $\psi_\Psi^\tout$ on a certain left ideal of $\B$.
We show in Theorem~\ref{convergence-of-asymptotic-functionals} below, that asymptotic functionals are non-zero
in theories of Wigner particles. If non-trivial asymptotic functionals arise in the absence of Wigner particles,
then we say that the theory describes infraparticles\footnote{The conventional definition of infraparticles requires
that both $\hil_+$ and $\hil_-$ contain at most multiples of the vacuum vector. Our (less restrictive) definition imposes  this
requirement on one of these subspaces only. Thus theories containing `waves' running to the right but no `waves' running to the left
(or vice versa) describe infraparticles according to our terminology. Such nomenclature turns out to be more convenient 
in the context of two-dimensional, massless theories.}. Using standard decomposition theory, 
the GNS representation 
$\pi$ induced by the sesquilinear form $\psi_{\Psi}^{\tout}$ can be decomposed into a direct integral of irreducible representations 
\beq
\pi\simeq \int_{X} d\mu(\xi)\,\pi_{\xi}, \label{decomposition-representation}
\eeq 
where $(X, d\mu)$ is a Borel space and $\simeq$ denotes unitary equivalence \cite{Ta,Po04.2}. Results from \cite{AH67,St89} suggest that the 
measurable field of irreducible representations $\{\pi_{\xi}\,\}_{\xi\in X}$ carries information about all the (infra-)particle types appearing in the theory. In particular, there exists a field of vectors $\{\, q_{\xi} \,\}_{\xi\in X}$  which can be interpreted as the
energy and momentum of plane wave configurations $\{\psi_{\xi}\}_{\xi\in X}$ of the respective (infra-)particles \cite{Po04.2}. 
The sesquilinear forms $\{\psi_{\xi}\}_{\xi\in X}$, called \emph{pure particle weights}, induce the
representations $\{\pi_{\xi}\}_{\xi\in X}$ and satisfy
\beq
\psi_{\Psi}^{\tout}=\int_X d\mu(\xi)\,\psi_{\xi}. \label{decomposition-first}
\eeq
The existence of such a decomposition was shown, under certain technical restrictions, in \cite{Po04.1,Po04.2}.

The theory of particle weights  is sufficiently general to accommodate the phenomenon of the infraparticle's momentum superselection, discussed above: In this case $q_{\xi}\neq q_{\xi'}$ should imply that 
$\pi_{\xi}$ is not unitarily equivalent to $\pi_{\xi'}$ for almost all labels $\xi$, $\xi'$ corresponding to the
infraparticle in question. Superselection of direction of motion is a milder property: It only 
requires that plane waves $\psi_{\xi}, \psi_{\xi'}$, travelling in opposite directions, give rise to representations $\pi_{\xi}, 
\pi_{\xi'}$ which are not unitarily equivalent. This latter interplay between the
infraparticle's kinematics and the superselection structure occurs in 
some two-dimensional conformal field theories, as we explain below. We state this property precisely in  
Definitions~\ref{superselection-weights} and  \ref{superselection}, where we restrict attention to representations $\pi$ of (Murray-von Neumann) type I with atomic center. This is sufficient for our purposes and allows us to separate our central concept from  ambiguities involved in the general decompositions~(\ref{decomposition-representation}), (\ref{decomposition-first}).

Our discussion of conformal field theory relies on the notion of a local net of von Neumann algebras on $\real$
which we introduce in Subsection~\ref{preliminaries-two}. (Such nets arise e.g. by restricting  the familiar
M\"obius covariant nets on the circle to the real line). With any open bounded region $\I\subset\real$ we associate
a von Neumann algebra $\A(\I)$, acting on a Hilbert space $\hilk$, and denote the
quasilocal algebra of this net by $\A$.
Moreover, the Hilbert space $\hilk$ carries a unitary representation of translations $\real\ni \t\to\U(\t)$, whose spectrum coincides with
$\real_+$. If there exists a  cyclic, unit vector $\Omz\in\hilk$, which is the unique (up to a phase) non-zero vector invariant under the action of $\U$,
then we say that the theory is in a vacuum representation. Given such a net, covariant under the action of some internal symmetry group, one can proceed to the fixed-point subnet which has a non-trivial superselection structure. In the simple case, considered in 
Subsection~\ref{infra},   the action of $\mathbb{Z}_2$  is implemented by a unitary $W\neq I$ on $\hilk$ s.t. $W^2=I$. The
fixed-point subnet $\A_{\ev}$ consists of all the elements of $\A$, which commute with $W$. The subspace
$\hilk_{\ev}=\ker(W-I)$  (resp. $\hilk_{\odd}=\ker(W+I)$) is invariant under the action of $\A_{\ev}$ and gives rise
to a vacuum representation (resp. a charged representation) of the fixed-point theory. 

Given two nets of von Neumann algebras on the real line, $\A_1$ and $\A_2$, acting on  Hilbert spaces $\hilk_1$ and $\hilk_2$, one
obtains the two-dimensional chiral net $\B$, acting on $\hil=\hilk_1\otimes\hilk_2$, by the standard 
construction, recalled in Subsection~\ref{preliminaries-two}: The two real lines are identified with the 
lightlines in $\real^2$ and for any double cone $\mco=\I\times \J$ one sets\footnote{In the main part of the paper $\B(\I\times\J)$ denotes a suitable weakly dense `regular subalgebra' of $\A_1(\I)\otimes \A_2(\J)$.  This distinction is not essential for the present introductory discussion.} 
$\B(\I\times\J)=\A_1(\I)\otimes \A_2(\J)$.
If the nets $\A_1$, $\A_2$ are in  vacuum representations, with the vacuum vectors $\Om_1\in\hilk_1$, $\Om_2\in\hilk_2$,
then $\B$ is also in a vacuum representation, with the vacuum vector $\Om=\Om_1\otimes\Om_2$. 
In spite of their simple tensor product structure, chiral nets play a prominent role
in conformal field theory. In fact, with any local conformal  net 
on $\real^2$ one can associate a chiral subnet by restricting the theory to the lightlines. 
In the important case of central charge $c<1$ these subnets were instrumental for
the classification results, mentioned above, which clarified the superselection structure of
a large class of models  \cite{Re00, KL04.1}. As we show in the present work,   
chiral nets  also offer a promising starting point for the analysis of  particle aspects of
conformal field theories: Any chiral net in a vacuum representation is an  asymptotically
complete theory of Wigner particles. Moreover, any charged irreducible product representation of
such a net contains infraparticles. With this information at hand, we exhibit examples of infraparticles with superselected direction of
motion. This construction is  summarized briefly in the remaining part of this Introduction.

Let us consider two fixed-point nets $\A_{1,\ev}$, $\A_{2,\ev}$, obtained from $\A_1$ and $\A_2$ with the help of
the unitaries $W_1$ and $W_2$, implementing the respective actions of  $\mathbb{Z}_2$. The resulting chiral
net $\B_{\ev}$ acts on the Hilbert space $\hil=\hilk_1\otimes\hilk_2$, which decomposes into four invariant subspaces
with different particle structure:
\beq
\hil=(\hilk_{1,\ev}\otimes\hilk_{2,\ev})\oplus(\hilk_{1,\odd}\otimes\hilk_{2,\ev})\oplus(\hilk_{1,\ev}\otimes\hilk_{2,\odd})\oplus(\hilk_{1,\odd}\otimes\hilk_{2,\odd}).
\eeq
$\B_{\ev}$ restricted to  $\hil_0:=\hilk_{1,\ev}\otimes\hilk_{2,\ev}$ is a chiral theory in a vacuum representation. Thus it is an asymptotically complete theory of Wigner particles, by the result mentioned above. $\hil_{\r}:=\hilk_{1,\odd}\otimes\hilk_{2,\ev}$ contains `waves' travelling to the right, but no `waves' travelling to the left. In $\hil_{\l}:=\hilk_{1,\ev}\otimes\hilk_{2,\odd}$ the opposite situation occurs. Thus $\B_{\ev}$ restricted to $\hil_{\r}$ or $\hil_{\l}$ describes infraparticles, according
to our terminology.  Finally, $\B_{\ev}$ restricted to  $\hiltt:=\hilk_{1,\odd}\otimes\hilk_{2,\odd}$ is a theory of infraparticles which does not contain `waves'. 
In Theorem~\ref{representation-decomposition-two} below, which is our main result, we establish superselection of direction of motion 
for infraparticles described by the net $\Btt=\B_{\ev}|_{\hiltt}$. The argument proceeds as follows: $\B_{\ev}$ is contained
in $\B$, which is an asymptotically complete theory of Wigner particles. Thus we can use the scattering theory from \cite{Bu75} 
to compute the asymptotic functionals~(\ref{asymptotic-functional-first}) and obtain the decompositions~(\ref{decomposition-representation}) of their GNS representations. Interpreted as a state on $\B$, any vector $\Psi_{1}\otimes\Psi_{2}\in\hiltt$ consists of two `waves'   at asymptotic times:
 $\Psi_1\otimes\Om_2$ travelling to the right and $\Om_1\otimes\Psi_2$ travelling to the left. 
  (Cf. Theorem~\ref{asymptotic-completeness-chiral} below). However, these  two vectors belong to  different invariant subspaces of $\B_{\ev}$, namely to  $\hil_{\r}$ and $\hil_{\l}$. 
The corresponding representations of $\Btt$ are not unitarily equivalent, since they have different structure of the energy-momentum spectrum.

Our paper is organized as follows: Section~\ref{scattering}, which does not rely on conformal symmetry, concerns  two-dimensional,
massless quantum field theories and their particle aspects: Preliminary Subsection~\ref{first-preliminaries} introduces
the main concepts. In  Subsection~\ref{Wigner} we recall the scattering theory of two-dimensional, massless Wigner particles 
developed in \cite{Bu75}. Subsection~\ref{weights} gives a brief exposition of the  theory of particle weights and introduces
our main concept: superselection of direction of motion. Subsection~\ref{asymptotic-functionals-sub} 
presents our main technical result,
stated in Theorem~\ref{convergence-of-asymptotic-functionals}, which clarifies the structure of asymptotic functionals in theories of Wigner particles. Its proof is given in Appendix~\ref{Appendix-A}. In Section~\ref{conformal-section} we apply the concepts and tools presented
in Section~\ref{scattering} to chiral conformal field theories. Our setting, which is slightly more general than the usual framework of  conformal field theory, is presented in Subsection~\ref{preliminaries-two}. In Subsection~\ref{scattering-two} we show that any chiral theory in a vacuum representation has a complete particle interpretation in terms of Wigner particles. In Subsection~\ref{infraparticles} we demonstrate that charged irreducible product representations of any chiral theory describe infraparticles.  Subsection~\ref{infra} presents our main result, that is  superselection of the infraparticle's direction of motion in chiral theories arising from fixed-point nets of $\mathbb{Z}_2$ actions. 
Proofs of some auxiliary lemmas are postponed to Appendix~\ref{Appendix-B}. In Section~\ref{conclusions} we summarize our work  and  discuss future directions.

\section{Particle aspects of two-dimensional massless theories}\label{scattering}
\setcounter{equation}{0}

\subsection{Preliminaries}\label{first-preliminaries}
In this section, which does not rely on  conformal symmetry, we present some general results 
on  particle aspects of massless quantum field theories in two-dimensional spacetime.
We rely on the following variant of the Haag-Kastler axioms \cite{Ha}:  
\begin{definition}\label{two-dim-net}
A local net of $C^*$-algebras on $\real^2$ is a pair $(\B,U)$ consisting of 
a map $\mco\to \B(\mco)$ from the family of open, bounded regions of $\real^2$ to the family of $C^*$-algebras on a Hilbert space 
$\hil$, and a strongly continuous  unitary representation of translations $\real^2\ni \x\to U(\x)$ acting on $\hil$, 
which are subject to the following conditions:
\begin{enumerate}
\item (isotony) If $\mco_1 \subset \mco_2$, then $\B(\mco_1)\subset \B(\mco_2)$.
\item (locality) If $\mco_1 \perp \mco_2$, then $[\B(\mco_1),\B(\mco_2)] = 0$, where $\perp$ denotes
spacelike separation.
\item (covariance)  $U(\x)\B(\mco)U(\x)^*=\B(\mco+\x)$ for any $\x\in\real^2$.
\item (positivity of energy) The spectrum of  $U$ is contained in the closed forward lightcone~$V_+:=\{\,(\om,\pb)\in\real^2\,|\,\om\geq |\pb|\,\}$. 
\item (regularity) The group of translation automorphisms $\al_x(\,\cdot\,)=U(\x)\,\cdot\,U(\x)^*$
satisfies $\lim_{\x\to 0}\|\al_{\x}(A)-A\|=0$ for any $A\in\B$.
\end{enumerate}
We also introduce the quasilocal $C^*$-algebra of this net  $\B=\ov{\bigcup_{\mco\subset\real^2}\B(\mco)}$.

\end{definition}
For any given net $(\B,U)$ there exists exactly one  unitary representation of 
translations $U^{\can}$ s.t. $U^{\can}$ implements $\al$, all the operators $U^{\can}(\x)$,
$\x\in\real^2$ are contained in $\B''$, the spectrum of $U^{\can}$ is contained 
in $V_+$  and has Lorentz invariant lower boundary upon restriction to any subspace of $\hil$ invariant under the
action of $\B''$ \cite{BB85}. We assume that this \emph{canonical} representation of translations has been selected above, 
i.e. $U=U^{\can}$. We denote by $(H,\Pb)$ the corresponding energy-momentum operators
i.e.  $U(\x)=e^{iHt-i\Pb\xb}$,  $\x=(t,\xb)$. As we are interested in scattering of massless particles,  we introduce
the single-particle subspaces $\H_{\pm}:=\ker(H\mp \Pb)$ and denote the corresponding projections by $P_{\pm}$.
The intersection $\hil_{+}\cap\hil_-$ contains only translationally invariant vectors. If $\hil_{+}\neq \hil_{+}\cap\hil_-$
and $\hil_-\neq  \hil_{+}\cap\hil_-$  then we say that the theory describes Wigner particles. If  $U$ has a 
unique (up to a phase) invariant unit vector $\Om\in\hil$ and $\Om$ is cyclic under the action of $\B$
then we say that the net $(\B,U)$ is in a vacuum representation. In this case $\B$ acts irreducibly on $\hil$.
(Cf. Theorem 4.6 of
\cite{Ar}). Scattering theory for Wigner particles in a vacuum representation, developed in \cite{Bu75}, will be
recalled in Subsection~\ref{Wigner}.

In the absence of Wigner particles we will apply the theory of particle weights \cite{Bu90, BPS91, Po04.1, Po04.2},
outlined in Subsection~\ref{weights}, to extract the (infra-)particle content of a given theory.
In this context it is necessary to consider various representations of the net $(\B,U)$.
A representation of the net $(\B,U)$ is, by definition, a family of representations
$\{\pi_\O\}$ of local algebras which are consistent in the sense that
if $\O_1 \subset \O_2$ then it holds that $\pi_{\O_2}|_{\B(\O_1)} = \pi_{\O_1}$.
Since the family of open bounded regions in $\RR^2$ is directed, this representation
uniquely extends to a representation $\pi$ of the quasilocal $C^*$-algebra $\B$.
Conversely, a representation of $\B$ induces a consistent family of representations of local
algebras. In the following $\pi$ may refer to a representation of $\B$ or a family
of representations. We say that
a representation $\pi :\B\to B(\hil_{\pi})$ is covariant, if there exists a strongly 
continuous group of unitaries $U_{\pi}$ on $\hil_{\pi}$, s.t. 
\beq
\pi(\al_{\x}(A))=U_{\pi}(\x)\pi(A)U_{\pi}(\x)^{*}, \quad A\in\B,\,\, \x\in\real^2. \label{automorphisms-zero}
\eeq
Moreover, we say that this representation has positive energy, if the joint spectrum of the
generators of $U_{\pi}$ is contained in $V_++q$ for some $q\in \real^2$. We denote the
corresponding canonical representation of translations by $U_{\pi}^{\can}$ and note that $(\pi(\B), U^{\can}_{\pi})$ is
again a local net of $C^*$-algebras in the sense of Definition~\ref{two-dim-net}. We say that the net $(\pi(\B), U^{\can}_{\pi})$  
is in a charged irreducible representation, if $\pi(\B)$ acts irreducibly on a non-trivial Hilbert space $\hil_{\pi}$ which does not contain non-zero invariant vectors of $U^{\can}_{\pi}$.

We call two representations $(\pi_1,\hil_{\pi_1})$ and $(\pi_2,\hil_{\pi_2})$ of $(\B,U)$ unitarily equivalent, 
(in short $(\pi_1,\hil_{\pi_1})\simeq (\pi_2,\hil_{\pi_2})$), if there exists a unitary $W:\hil_{\pi_1}\to\hil_{\pi_2}$
s.t. 
\beqa
W\pi_1(A)=\pi_2(A)W, \quad \ph{444} A\in\B.   \label{equivalence-one}
\eeqa 
If $\pi_1$ is a covariant, positive energy representation then so is $\pi_2$ and it is easy to see that 
\beq
WU_{\pi_1}^{\can}(\x)=U_{\pi_2}^{\can}(\x)W, \quad \x\in \real^2. \label{equivalence-two}
\eeq

\begin{remark}

We note that our (non-standard) Definition~\ref{two-dim-net} of the local net  neither
imposes the Poincar\'e covariance nor the existence of the vacuum vector. Thus it 
applies both to vacuum representations and charged representations, 
which facilitates our discussion. Apart from the physically motivated assumptions,
we adopt the regularity property~5, which can always be assured at the cost
of proceeding to a weakly dense subnet. This property seems indispensable
in the general theory of particle weights \cite{Po04.1} e.g. in the proof of 
Proposition~\ref{regularity-proposition} stated  below. For consistency
of the presentation, we proceed to regular subnets also in our discussion
of conformal field theories in Section~\ref{conformal-section}. We stress, however, that this property is
not needed there at the technical level.

\end{remark}

\subsection{Scattering states}\label{Wigner}

Scattering theory for Wigner particles in a vacuum representation of a two-dimensional massless theory $(\B,U)$ 
was developed in \cite{Bu75}. For the reader's convenience we recall here the main steps of this
construction.  Following \cite{Bu75}, for any $F\in\B$ and $T\geq 1$ we introduce the approximants:
\beqa
F_\pm(h_T)= \int h_T(t) F(t, \pm t) dt,
\eeqa
where $F(\x):=\al_{\x}(F)$, $h_T(t)=|T|^{-\eps}h(|T|^{-\eps}(t-T))$, $0<\eps<1$ and $h\in C_0^{\infty}(\real)$ is a non-negative function
s.t. $\int dt\, h(t)=1$. By applying  the mean ergodic theorem, one obtains
\beq
\lim_{T\to\infty}F_\pm(h_T)\Om=P_{\pm}F\Om. \label{ergodic-theorem}
\eeq
Moreover, for $F\in \B(\mco)$ and sufficiently large $T$ the operator  $F_{+}(h_T)$ (resp. $F_-(h_T)$)
commutes with any observable localized in the left (resp. right) component of the spacelike complement of $\mco$. 
Exploiting these two facts, the following result was established in \cite{Bu75}: 
\bep[\cite{Bu75}]\label{scattering-first-lemma} Let $F, G\in \B$.  Then the limits
\beqa
\Phi_{\pm}^{\tout}(F):=\underset{T\to\infty}{\slim}\; F_{\pm}(h_T) \quad \label{asymptotic-field}
\eeqa
exist and are called the (outgoing) asymptotic fields. They depend only on the respective vectors $\Phi_{\pm}^{\tout}(F)\Om=P_{\pm}F\Om$ and  satisfy:
\begin{enumerate}
\item[(a)] $\Phi_+^{\tout}(F)\hil_+\subset\hil_+, \quad\Phi_-^{\tout}(G)\hil_-\subset\hil_-$.

\item[(b)] $\al_{\x}(\Phi_+^{\tout}(F))=\Phi_+^{\tout}(\al_{\x}(F)),\quad \al_{\x}(\Phi_-^{\tout}(G))=\Phi_-^{\tout}(\al_{\x}(G))$ for $\x\in\RR^2$.

\item[(c)] $[\Phi_+^{\tout}(F), \Phi_-^{\tout}(G)]=0$.

\end{enumerate}
The incoming asymptotic fields $\Phi_{\pm}^{\tin}(F)$ are constructed analogously, by taking
the limit $T\to-\infty$.
\eep
\noindent With the help of the asymptotic fields one defines the scattering states as follows: Since $\B$ acts irreducibly
on $\hil$, for any $\Psi_{\pm}\in\hil_{\pm}$ we can find $F_{\pm}\in\B$ s.t. $\Psi_{\pm}=F_{\pm}\Om$ \cite{Sa}. The vectors
\beqa
\Psi_+\pout\Psi_-=\Phi_{+}^{\tout}(F_+)\Phi_{-}^{\tout}(F_-)\Om
\eeqa
are called the (outgoing) scattering states. By Proposition~\ref{scattering-first-lemma} they do not depend on the choice of
$F_{\pm}$ within the above restrictions. The incoming scattering states $\Psi_+\pin\Psi_-$ are defined analogously.
The physical interpretation of these vectors, as two independent excitations travelling in opposite directions
at asymptotic times, relies on the following proposition from \cite{Bu75}:
\bep[\cite{Bu75}]\label{scattering-second-lemma} Let $\Psi_{\pm},\Psi_{\pm}'\in\hil_{\pm}$. Then:
\begin{enumerate}
\item[(a)] $(\Psi_+\pout \Psi_-|\Psi'_+\pout \Psi'_-)=(\Psi_+|\Psi'_+)(\Psi_-|\Psi'_-)$,
\item[(b)] $U(\x)(\Psi_+\pout \Psi_-)=(U(\x)\Psi_+)\pout (U(\x)\Psi_-)$, for $\x\in\real^2$.
\end{enumerate}
Analogous relations hold for  the incoming scattering states. 
\eep
\noindent Following \cite{Bu75}, we define the  subspaces spanned by the respective scattering states:
\beq
\hil^{\tin}=\hil_+\pin\hil_-\,\, \textrm{ and }\,\, \hil^{\tout}=\hil_+\pout \hil_-.
\eeq
Next, we introduce the wave operators
$\Om^{\tout}:\hil_+\otimes \hil_-\to \hil^{\tout}$ and $\Om^{\tin}:\hil_+\otimes \hil_-\to \hil^{\tin}$,
extending by linearity the relations
\beq
\Om^{\tout}(\Psi_+\otimes\Psi_-)=\Psi_+\pout \Psi_- \,\,\textrm{ and }\,\, \Om^{\tin}(\Psi_+\otimes\Psi_-)=\Psi_+\pin \Psi_-.
\eeq
These operators are isometric in view of Proposition~\ref{scattering-second-lemma} (a).
The scattering operator $S: \hil^{\tout}\to\hil^{\tin}$, given by
\beq
S=\Om^{\tin}(\Om^{\tout})^*,   \label{scattering-matrix}
\eeq
is also an isometry. Now we are ready to  introduce two important concepts:
\bed (a) If  $S=I$  on $\hil^{\tout}$, then we say that the theory is non-interacting.\\
(b) If $\hil^{\tin}=\hil^{\tout}=\hil$ then we say that the theory is asymptotically complete
(in terms of `waves').
\eed
\noindent We  show in Theorem~\ref{asymptotic-completeness-chiral} below  that any chiral conformal field theory in a vacuum
representation is both non-interacting  and asymptotically complete. (We demonstrated these facts already in
\cite{DT10} in a different context).

To conclude this subsection, we introduce some other useful concepts which are needed in 
Theorem~\ref{convergence-of-asymptotic-functionals} below: Let us choose some closed subspaces $\K_{\pm}\subset\hil_{\pm}$,
invariant under the action of $U$,
and denote by $\K_{+}\pout \K_{-}$ the linear span of the respective scattering states.
For any $\Psi\in\K_{+}\pout \K_{-}\subset \hil^{\tout}$ we introduce the positive functionals $\rho_{\pm,\Psi}$, 
given by the relations
\beqa
\rho_{+,\Psi}(A)&=&((\Om^{\tout})^{-1}\Psi|(A\otimes I)(\Om^{\tout})^{-1}\Psi), \label{density+}\\ 
\rho_{-,\Psi}(A)&=&((\Om^{\tout})^{-1}\Psi|(I\otimes A)(\Om^{\tout})^{-1}\Psi), \label{density-}    
\eeqa 
where $A\in B(\hil)$ and the embedding $\K_+\otimes\K_-\subset \hil\otimes\hil$ is understood.
These functionals can be expressed as follows
\beq 
\rho_{\pm, \Psi}(\,\cdot\,)=\sum_{n\in\nat}(\Psi_{\pm,n}|\,\cdot\, \Psi_{\pm,n}), \label{reduced-matrix}
\eeq
where $\Psi_{\pm,n}\in\K_{\pm}$ and $\sum_{n\in\nat} \|\Psi_{\pm,n}\|^2=\|\Psi\|^2$. 
It follows easily from Lemma~\ref{bases}, that for $\Psi\in P_E(\K_+\pout \K_-)$, where $P_E$
is the spectral projection on vectors of energy not larger than $E$, one can choose $\Psi_{\pm,n}\in P_E \K_\pm$. 
We note that for $\|\Psi\|=1$ the functionals $\rho_{\pm,\Psi}$  are just the familiar reduced density matrices. 

\subsection{Particle weights} \label{weights}

Similarly as in the previous subsection we consider a local net of $C^*$-algebras $(\B,U)$ acting on a Hilbert space
$\hil$. However, we do not assume that $\hil$ contains the vacuum vector or non-trivial single-particle subspaces
$\hil_{\pm}$.  To study  particle aspects  in this general situation we  use the theory of particle weights \cite{Bu90, BPS91, Po04.1, Po04.2}  which we recall in this and the next  subsection. With the help of this theory we formulate in  
Definitions~\ref{superselection-weights} and \ref{superselection} below
the central notion of this paper: superselection of direction of motion.

First, we recall two  useful concepts: almost locality and the energy decreasing property.   
An observable $B\in \B$ is called almost local, 
if there exists a net of operators $\{\, B_r\in \B(\mco_r)\,|\, r>0\,\}$, s.t. for any $k\in\nat_0$
\beq
\lim_{r\to\infty} r^k\|B-B_r\|=0,
\eeq
where $\mco_r=\{(t,\xb)\in\real^2\,|\, |t|+|\xb|<r\,\}$. We say that an operator $B\in \B$ is energy decreasing, 
if its energy-momentum transfer is a compact set which does not intersect with the closed forward lightcone $V_+$. We recall
that the energy-momentum transfer (or the Arveson spectrum w.r.t. $\al$) of an observable $B\in \B$ is
the closure of the union of supports of the distributions
\beq
(\Psi_1|\ti B(p)\Psi_2)=(2\pi)^{-1}\int d^2 x\,e^{-ip\x} (\Psi_1|B(\x)\Psi_2) \label{energy-momentum-transfer}
\eeq
over all $\Psi_1,\Psi_2\in\hil$, where $p=(\om,\pb)$, $\x=(t,\xb)$ and $p\x=\om t-\pb\xb$.

Following \cite{BPS91,Po04.1}, we introduce the subspace $\mcL_0\subset \B$, spanned by operators which are both 
almost local and energy decreasing, and  the corresponding left ideal in $\B$: 
\beq
\mcL:=\{\, AB\,|\, A\in \B, B\in\mcL_0\,\}. \label{left-ideal}
\eeq 
Particle weights form a specific class of sesquilinear forms on $\mcL$:
\begin{definition}\label{weight-def}
A particle weight is a non-zero, positive sesquilinear form $\psi$ on the left ideal
$\mcL$, satisfying the following conditions:
\begin{enumerate}
\item For any $L_1, L_2 \in \mcL$ and $A \in \B$ the relation $\psi(AL_1,L_2) = \psi(L_1,A^*L_2)$ holds.
\item For any $L_1,L_2 \in \mcL$ and $\x\in\real^2$ the relation  $\psi(\al_{\x}(L_1),\al_{\x}(L_2)) = \psi(L_1,L_2)$ holds.
\item For any $L_1, L_2 \in \mcL$ the map $\real^2\ni \x\to  \psi( L_1,\al_{\x}(L_2))$ is continuous. Its
Fourier transform is supported in a shifted lightcone $V_+-q$, where  $q\in V_+$ does not depend on $L_1,L_2$.
\end{enumerate}
\end{definition}
Let us now summarize the pertinent properties of particle weights established in \cite{Po04.1} (in a slightly
different framework). As a consequence of Theorem~\ref{integrability}, stated below,  particle weights satisfy 
the following clustering property \cite{Po04.1}
\beq
\int d\xb\,|\psi(L_1, \al_{\xb}(L_2))|<\infty,
\eeq
valid for $L_1=B_1^*A_1B_1'$, $L_2=B_2^*A_2B_2'$, where $B_1,B_1',B_2,B_2'\in \mcL_0$ and $A_1,A_2\in \B$ are almost local.
In view of this bound, the GNS representation  $(\pi_\psi, \hil_{\pi_\psi})$ induced by a particle weight $\psi$ is  
well suited for a description of physical systems which are  localized in space (e.g. configurations of particles). 
The Hilbert space $\hil_{\pi_\psi}$ is given by 
\beq
\hil_{\pi_\psi}=(\,\mcL/\{\,L\in\mcL\,|\, \psi(L,L)=0 \})^{\te{cpl}} \label{GNS-one}
\eeq
and the respective equivalence class of an element $L\in\mcL$ is denoted by $|L\ran\in\hil_{\pi_\psi}$. The completion is taken w.r.t. the scalar product $\lan L_1|L_2\ran:=\psi(L_1, L_2)$. The representation $\pi_{\psi}$ acts on $\hil_{\pi_\psi}$ as follows
\beq
\pi_{\psi}(A)|L\ran=|AL\ran,\quad A\in \B.\label{GNS-two}
\eeq  
This representation is covariant and the translation automorphisms are implemented by the strongly continuous group 
of unitaries $U_{\pi_\psi}$, given by
\beq
U_{\pi_\psi}(\x)|L\ran=|\al_{\x}(L)\ran,\quad \x\in\real^2,\, L\in\mcL \label{standard}
\eeq        
which is called the standard representation of translations in the representation $\pi_{\psi}$.
By property 3 in Definition~\ref{weight-def} above, its spectrum is contained in a shifted closed forward lightcone. The corresponding
canonical representation  will be denoted by $U_{\pi_\psi}^\can$. (Cf. the discussion below  Definition~\ref{two-dim-net}). We also introduce  operators $(Q^0, \bQ)$ of \emph{characteristic energy-momentum} of $\psi$ which are the generators of the following group of unitaries on $\hil_{\pi_\psi}$
\beq
\V_{\pi_\psi}(x)=U^\can_{\pi_\psi}(x)U_{\pi_\psi}(x)^{-1}\in \pi_{\psi}(\B)', \label{characteristic-operators}
\eeq
i.e. $\V_{\pi_\psi}(x)=e^{iQ^0t-i\bQ\xb}$. 
We call a particle weight \emph{pure}, if its GNS representation is irreducible. It follows from definition~(\ref{characteristic-operators}) that the operator of characteristic energy-momentum of such a weight is a vector $q=(q^0,\bq)\in \real^2$. It can be interpreted as the energy and momentum of the plane wave configuration of the particle  described by this weight \cite{AH67,Po04.1}.

To extract properties of elementary  subsystems (particles) of a physical system described by a given (possibly non-pure) particle weight, it is natural to study irreducible subrepresentations of its GNS representation.
To ensure that there are sufficiently many such subrepresentations, we restrict attention to 
particle weights $\psi$ whose GNS representations $\pi_{\psi}$ are
of type I with atomic center\footnote{i.e. whose center is a direct sum of one-dimensional von Neumann algebras.}. 
(In particular, $\pi_\psi$ appearing in our examples in Subsection~\ref{infra} below belong to this family).
Then, by Theorem 1.31 from Chapter V of \cite{Ta}, there exists a unique family of Hilbert spaces $(\hh_{\al},\kk_{\al})_{\al\in \II}$  
and a unitary $W:\hil_{\pi_{\psi}}\to \bigoplus_{\al\in\II}\{\hh_{\al}\otimes\kk_{\al}\}$ s.t.
\beqa
W\pi_{\psi}(\B)''W^{-1}&=&\bigoplus_{\al\in\II}\{ B(\hh_{\al})\otimes \CC I\}, \label{double-commutant}  \\  
W\pi_{\psi}(\B)'W^{-1}&=&\bigoplus_{\al\in\II}\{ \CC I\otimes B(\kk_{\al}) \}. \label{commutant} 
\eeqa
We note that a subspace $\K_{\al,e}\subset\hil_{\pi_{\psi}}$ carries an irreducible 
subrepresentation $\pi_{\al,e}$ of $\pi_{\psi}$,  if and only if $W\K_{\al,e}=\hh_{\al}\otimes \CC e$ 
for some $\al\in\II$ and $e\in\kk_{\al}$. Clearly, $\pi_{\al,e}$ and $\pi_{\al,e'}$ are unitarily 
equivalent for any fixed $\al$ and arbitrary vectors $e, e'\in\kk_{\al}$. Choosing in any $\kk_{\al}$ an
orthonormal basis $B_{\al}$, we obtain
\beq
\pi_{\psi}=\bigoplus_{\substack{\al\in\II \\ e\in B_{\al}} }\pi_{\al,e}. \label{direct-sum-decomposition}
\eeq 
It is clear from the above discussion that any irreducible subrepresentation of $\pi_{\psi}$ is unitarily
equivalent to some $\pi_{\al,e}$ in the decomposition above.

If all the representations in the decomposition~(\ref{direct-sum-decomposition}) are unitarily equivalent
to some fixed vacuum representation, then we call the particle weight $\psi$ neutral. Otherwise we call 
$\psi$ charged. In the case of charged particle weights there may occur an interplay between the
translational and internal degrees of freedom of the system  which we call \emph{superselection of direction
of motion}. To introduce this concept, we need some terminology: Let $\hil_{\pi_\psi,\rr}$ (resp. $\hil_{\pi_\psi,\ll}$) 
be the spectral subspace of the characteristic momentum operator $\bQ$ of $\psi$,  
corresponding to the interval $[0,\infty)$ (resp. $(-\infty,0)$). 
Let $\pi$ be an irreducible subrepresentation  of $\pi_{\psi}$,
acting on a subspace $\K\subset\hil_{\pi_\psi}$. Then we say that $\pi$ is \emph{right-moving} (resp. \emph{left-moving}),
if $\K\neq\{0\}$ and $\K\subset\hil_{\pi_\psi,\rr}$  (resp. $\K\subset\hil_{\pi_\psi,\ll}$). By a suitable choice of the 
bases $B_{\al}$ one can ensure that each representation $\pi_{\al,e}$, appearing in  decomposition~(\ref{direct-sum-decomposition}), has one of these properties. (In fact, exploiting relations~(\ref{characteristic-operators}), (\ref{commutant}), one can choose such basis vectors $e\in\kk_{\al}$ that $W^{-1}(\hh_{\al}\otimes \CC e)$ belong  to $\hil_{\pi_\psi,\rr}$ or 
$\hil_{\pi_{\psi},\ll}$). 
After this preparation we define the central concept of the present paper:
\bed\label{superselection-weights} Let $\mcW$ be a family of particle weights and assume that their GNS 
representations $\{\,(\pi_{\psi}, \hil_{\pi_\psi}) \,|\, \psi \in\mcW\,\}$ are of type I with atomic centers. 
Suppose that for any $\psi,\psi'\in\mcW$ the following properties hold:
\begin{enumerate}
\item  $\pi_{\psi}$ has both left-moving and right-moving irreducible subrepresentations. 
\item  No right-moving, irreducible 
subrepresentation of $\pi_{\psi}$ is unitarily equivalent to a left-moving irreducible 
subrepresentation of $\pi_{\psi'}$.
\end{enumerate}
Then we say that this family of particle weights has superselected direction of motion.
\eed

Let us now relate superselection of direction of motion in the above sense to our discussion of this concept in the Introduction.  
For this purpose we consider a particle weight $\psi$, whose GNS representation is of type I with atomic center and acts on
a \emph{separable} Hilbert space $\hil_{\pi_{\psi}}$.    
 Making use of formula~(\ref{direct-sum-decomposition}) and identifying unitarily each $\pi_{\al,e}$, acting on $W^{-1}(\hh_{\al}\otimes \CC e)$, with $\pi_{\al}:=\pi_{\al,e_0}$ acting on $\K_{\al}:=W^{-1}(\hh_{\al}\otimes \CC e_0)$ 
for some chosen $e_0\in B_{\al}$,  we obtain
\beq
\pi_{\psi}(\B)\simeq \bigoplus_{\al\in\II}\{\pi_{\al}(\B)\otimes \CC I\}, \label{direct-sum}
\eeq
where the r.h.s. acts on $\bigoplus_{\al\in\II}\{ \K_{\al}\otimes \kk_{\al}\}$. In the sense of the
same identification
\beq
\pi_{\psi}(\B)'\simeq \bigoplus_{\al\in\II}\{\CC I \otimes B(\kk_{\al})\}. \label{commutant1}
\eeq
Now, following \cite{Po04.2}, we choose a  maximal abelian von Neumann algebra $\M$   in  $\pi_{\psi}(\B)'$, 
containing $\{\,\V_{\pi_{\psi}}(x)\,|\, x\in\real^2\, \}$. As a consequence of formula~(\ref{commutant1})  
\beq
\M \simeq \bigoplus_{\al\in\II}\{ \CC I\otimes \M_{\al} \},
\eeq
where $\M_{\al}\subset B(\kk_{\al})$ are maximal abelian von Neumann subalgebras. 
For any such $\M_{\al}$ there exists a Borel space $(Z_{\al},d\mu_{\al})$ 
s.t. $(\M_{\al},\kk_{\al})\simeq (L^{\infty}(Z_{\al},d\mu_{\al}), L^{2}(Z_{\al},d\mu_{\al}) )$. (This fact uses 
separability of the Hilbert space. See Theorem~II.2.2 of \cite{Da}). 
Adopting this identification in (\ref{direct-sum}) and (\ref{commutant1}), we obtain 
\beq
\V_{\pi_{\psi}}\simeq \bigoplus_{\al\in\II}\{ I\otimes \V_{\al} \}, \label{Vsimeq}
\eeq 
where $\V_{\al}(x)\in L^{\infty}(Z_{\al},d\mu_{\al})$ is the operator of multiplication by  (the equivalence class of) the function 
$Z_{\al}\ni z\to e^{iq_{\al,z}x}$, where $q_{\al,z}=(q_{\al,z}^0, \bq_{\al,z} )\in \real^2$.
Introducing the field of representations $(\pi_{\al,z},\hh_{\al,z})_{z\in Z_{\al}}$ s.t.
$\pi_{\al,z}=\pi_{\al}$ and $\hh_{\al,z}=\K_{\al}$ for all $z\in Z_{\al}$, we obtain from relation~(\ref{direct-sum})
the existence of a unitary $\ti W:\hil_{\pi_\psi}\to \bigoplus_{\al\in\II}\int^{\oplus}d\mu_{\al}(z)\,\hh_{\al,z}$
s.t.
\beq
\ti W\pi_{\psi}(\,\cdot\,)\ti W^{-1}= \bigoplus_{\al\in\II}\int^{\oplus}_{Z_{\al}}d\mu_{\al}(z)\,\pi_{\al,z}(\,\cdot\,).\label{direct-int}
\eeq
This is an example of decomposition~(\ref{decomposition-representation}), stated in the Introduction.
Moreover, as a consequence of (\ref{Vsimeq}), 
\beq
\ti W \V_{\pi_{\psi}}(x) \ti W^{-1}= \bigoplus_{\al\in\II}\int^{\oplus}_{Z_{\al}}d\mu_{\al}(z)\,e^{iq_{\al,z}x},
\eeq
where $\{q_{\al,z}\}_{z\in Z_{\al}}$ is the field of characteristic energy-momentum vectors\footnote{This terminology is
consistent with the discussion after formula~(\ref{characteristic-operators}). In fact, under some technical
restrictions each  $\pi_{\al,z}$ is induced by
some pure particle weight $\psi_{\al,z}$, whose characteristic energy-momentum vector is $q_{\al,z}$ \cite{Po04.1,Po04.2}. 
Cf. also formula~(\ref{decomposition-first}).  }
of the representations 
$( \pi_{\al,z}, \hh_{\al,z} )_{z\in Z_{\al}}$. As we required in the Introduction, for any particle weight with superselected
direction of motion, the relation $\bq_{\al,z}\cdot \bq_{\al',z'}\leq 0$ should imply that $\pi_{\al,z}$ is not unitarily equivalent to 
$\pi_{\al',z'}$ for almost all $z,z'$. This is in fact the case in view of the following proposition.     
\bep Suppose that $\psi$ belongs to a family of particle weights which has superselected direction of motion in the
sense of Definition~\ref{superselection-weights} and s.t. its GNS representation acts on a separable Hilbert space. Then $\{\pi_{\al,z}\}_{z\in Z_{\al}}$, appearing in the decomposition (\ref{direct-int}) of $\pi_{\psi}$, 
is a field of right-moving (resp. left-moving) representations, if and only if $\bq_{\al,z}\geq 0$ 
(resp. $\bq_{\al,z}<0$) for almost all $z\in Z_{\al}$.
\eep
\proof Suppose that $\pi_{\al}$ is a right-moving subrepresentation of $\pi_{\psi}$ i.e. $\K_{\al}\subset\hil_{\pi_\psi,\rr}$.
We recall that $\pi_{\al}$ coincides with $\pi_{\al,e_0}$ acting on $\K_{\al,e_0}=\K_{\al}$.
Since every $\pi_{\al,e}$, $e\in \kk$, is unitarily equivalent to  $\pi_{\al,e_0}$,   
the property of superselection of  direction of motion implies that $\K_{\al,e}\subset\hil_{\pi_\psi,\rr}$ 
for all $e\in\kk$. Consequently,  $\hil_{\al}=W^{-1}(\hh_{\al}\otimes\kk_{\al})\subset \hil_{\pi_\psi,\rr}$. Since
the projection $P_{\al}$ on $\hil_{\al}$ is central, this subspace is invariant under the action of $\V_{\pi_{\psi}}$.
Formula~(\ref{Vsimeq}) gives
\beq
\V_{\pi_{\psi}}(x)P_{\al}\simeq I\otimes \V_{\al}(x), \label{VQ}
\eeq  
thus the spectra of the generators of $\real^2\ni x\to\V_{\al}(x)$ and $\real^2\ni x\to\V_{\pi_{\psi}}(x)P_{\al}$ coincide. 
In particular the spectrum of the generator of space translations of $\V_{\al}$ is contained in $[0,\infty)$.
The opposite implication follows immediately from relation~(\ref{VQ}). \qed

\subsection{Asymptotic functionals}\label{asymptotic-functionals-sub}

In this subsection we consider a concrete class of particle weights, introduced  in \cite{Bu87,Bu90, Po04.1}, which have applications in  scattering theory. Their construction relies on the following result due to  Buchholz (which remains valid in higher dimensions).
\bet[\cite{Bu90}]\label{integrability} Let $(\B,U)$ be a local net of $C^*$-algebras on $\real^2$.
Then, for any $E\geq 0$, $L\in\mcL$,
\beq
\left\|P_E\int_K d\xb\, (L^*L)(\xb) P_E\right\|\leq c,
\eeq 
where $P_E$ is the spectral projection on vectors of energy not larger than $E$, $K\subset \real$ is a compact interval, and $c$ is a  constant independent of~$K$.
\eet
\noindent Following \cite{Po04.1}, we introduce the algebra of detectors
$\mcC= \mathrm{span}\{L_1^*L_2: L_1, L_2 \in \mcL\}$ and 
equip it with a locally convex topology, given by the family of seminorms
\beq
p_E(C)=\sup \left\{ \int d\xb\, |(\Psi|C(\xb)\Psi)| \,\,|\,\,  \Psi\in P_E\hil, \,\|\Psi\|\leq 1\, \right\},\,  \quad C\in\mcC, 
\eeq
labelled by $E\geq 0$, which are finite by Theorem~\ref{integrability}.
Next, for any $\Psi\in\hil$ of bounded energy, (i.e. belonging to $P_E\hil$ for some $E\geq 0$),  we define  a sequence of functionals $\{\si_{\Psi}^{(T)}\}_{T\in\real}$ from the topological dual of $\mcC$: 
\beq
\si_{\Psi}^{(T)}(C):=\int dt\, h_T(t)\int d\xb\, (\Psi|C(t,\xb)\Psi),\quad  C\in\mcC. \label{approximants}
\eeq
As this sequence is uniformly bounded in $T$ w.r.t. any seminorm $p_E$,  the Alaoglu-Bourbaki
theorem gives limit points  $\si_{\Psi}^{\tout}\in\mcC^*$ as $T\to\infty$,  which are called the asymptotic functionals.
The following fact  was shown in \cite{Po04.1}: 
\bep[\cite{Po04.1}]\label{regularity-proposition} If $\si_{\Psi}^{\tout}\neq 0$, then the sesquilinear forms on $\mcL$, given by
\beq
\psi_{\Psi}^{\tout}(L_1,L_2):=\si_{\Psi}^{\tout}(L_1^*L_2),
\eeq
are particle weights, in the sense of Definition~\ref{weight-def}. 
\eep
 Fundamental results from \cite{AH67} suggest a physical interpretation of the particle weights $\psi_{\Psi}^{\tout}$ 
as mixtures of plane wave configurations of all the particle types described by the theory. 
(Cf. formulas~(\ref{decomposition-representation}), (\ref{decomposition-first})). 
Accordingly, we  say that a given theory has a non-trivial \emph{particle content}, 
if it admits some non-zero asymptotic functionals $\si_{\Psi}^{\tout}$.
This is the case in any massless two-dimensional theory of  Wigner particles (in a vacuum representation) 
as a consequence of the following theorem. A proof of this statement, which is our main technical result, 
is given in Appendix~\ref{Appendix-A}.
\bet\label{convergence-of-asymptotic-functionals} Let  $(\B,U)$ be a local net of $C^*$-algebras on $\real^2$ in a vacuum representation,
acting on a Hilbert space $\hil$.
Then, for any $\Psi\in P_E\hil^{\tout}$, $E\geq 0$,
\beqa
\psi_{\Psi}^{\tout}(L_1,L_2)&=&\lim_{T\to\infty}\int dt\, h_T(t)\int d\xb\, (\Psi|(L_1^*L_2)(t,\xb)\Psi) \non\\
&=&\int d\xb\,(\rho_{+,\Psi}+\rho_{-,\Psi})\big((L_1^*L_2)(\xb)\big), \label{main-technical}
\eeqa
where the functionals $\rho_{\pm,\Psi}$ are defined by (\ref{density+}), (\ref{density-}).
In particular, $\psi_{\Psi}^{\tout}=0$, if and only if $\Psi\in \CC\Om$. 
\eet
\noindent In a theory of Wigner particles  $\CC\Om\neq \hil_{\pm}\subset \hil^{\tout}$, 
thus the particle content is non-trivial by the above result. However, non-zero asymptotic functionals may also appear in 
the absence of Wigner particles i.e. when one or both of the subspaces $\hil_{\pm}$ equal $\hil_+\cap\hil_-$.
If this is the case, then we say that the net $(\B,U)$ describes \emph{infraparticles}. Theorem~\ref{existence-of-particles}  
below provides  a large class of such theories. In Theorem~\ref{representation-decomposition-two} we show that some of
these models describe excitations whose direction of motion is superselected in the following sense: 
\bed\label{superselection} Let $(\B,U)$ be a net  describing infraparticles. We say that the infraparticles
of the net $(\B,U)$ have superselected direction of motion, if $\{\, \psi_{\Psi}^{\tout}\,|\, \Psi\neq 0,\, 
\Psi\in P_E\hil, \,  E\geq 0\, \}$ is a family of particle weights with superselected direction of motion in the sense of 
Definition~\ref{superselection-weights}.    
\eed


\section{Particle aspects of conformal field theories}\label{conformal-section}

\setcounter{equation}{0}
\subsection{Preliminaries}\label{preliminaries-two}

In this section we are interested in particle aspects of chiral conformal field theories.
To emphasize the relevant properties of these models, we base our investigation
on the concept of a local net of von Neumann algebras on $\RR$, defined below. 
There are many examples of such nets. In particular, they arise from M\"obius covariant nets on $S^1$ by means of the
Cayley transform and the subsequent restriction to the real line. 
The simplest example is the so-called $U(1)$-current net \cite{BMT88}, whose subnets and extensions are well-studied.
For certain classes of nets on $S^1$ even classification results have been obtained \cite{KL04.1,KL04.2}.
\begin{definition}\label{circle-theory}
A local net of von Neumann algebras on $\real$ is a pair $(\A,\U)$ consisting of 
a map $\I\to \A(\I)$ from the family of open, bounded subsets of $\real$ to the
family of von Neumann
algebras on a Hilbert space $\hilk$ 
and a strongly continuous  unitary representation of translations $\real\ni \t\to \U(\t)$, acting on $\hilk$, 
which are subject to the following conditions:
\begin{enumerate}
\item (isotony) If $\I \subset \J$, then $\A(\I)\subset \A(\J)$.
\item (locality) If $\I \cap \J = \emptyset$, then $[\A(\I),\A(\J)] = 0$.
\item (covariance) $\U(\t)\A(\I)\U(\t)^* = \A(\I+\t)$ for any $\t\in\real$.
\item (positivity of energy) The spectrum of $\U$ coincides with $\real_+$.
\end{enumerate}
We also denote by $\A$ the quasilocal $C^*$-algebra of this net i.e. $\A=\ov{\bigcup_{\I\subset\real}\A(\I)}$. 
\end{definition}
Since we assumed that $\A(\I)$ are von Neumann algebras, we cannot demand norm
continuity of the functions $\t\to\alb_{\t}(A)$, $A\in\A$, where $\alb_{\t}(\,\cdot\,)=\U(\t)\,\cdot\,\U(\t)^*$. 
This regularity property holds, however, on the following weakly dense subnet of $C^*$-algebras   
\beq
\I\to\hA(\I):=\{\, A\in\A(\I)\, |\, \lim_{\t\to 0}\|\alb_{\t}(A)-A\|=0\, \}.
\eeq
The corresponding quasilocal algebra is denoted by $\hA$.

If  $\U$ has a unique (up to a phase) invariant, unit vector $\Omz\in\hilk$
and $\Omz$ is cyclic under the action of any $\A(\I)$ (the Reeh-Schlieder property) then we say that the net $(\A,\U)$ is in 
a vacuum representation. In this case $\A$ acts irreducibly on $\hilk$.
In the course of our analysis we will also consider other representations of  $(\A,\U)$. We say that
a representation $\pi :\A\to B(\hilk_{\pi})$ is covariant, if there exists a strongly 
continuous group of unitaries $\U_{\pi}$ on $\hilk_{\pi}$, s.t. 
\beq
\pi(\al_{\t}(A))=\U_{\pi}(\t)\pi(A)\U_{\pi}(\t)^{*}, \quad A\in\A,\,\, \t\in\real. \label{automorphisms-one}
\eeq
Moreover, we say that this representation has positive energy, if the spectrum of 
$\U_{\pi}$ coincides with $\real_+$. If $\pi$ is locally normal (i.e. its restriction to any local algebra $\A(\I)$
is normal) then  $(\pi(\A),\U_{\pi})$ is again a net of von Neumann algebras in the sense of Definition~\ref{circle-theory}.

Let $(\A_1,\U_1)$ and $(\A_2,\U_2)$ be two nets of von Neumann algebras on $\RR$,  acting on Hilbert spaces
$\hilk_1$ and $\hilk_2$. To construct a local net $(\B, U)$ on $\RR^2$, acting on the tensor product space $\H = \K_1\otimes \K_2$,  
we identify the two real lines with the lightlines $I_{\pm}=\{\,(t,\xb)\in\real^2\,|\, \xb\mp t=0\,\}$ in $\RR^2$. 
We first specify the unitary representation of translations
\beq
U(t,\xb):= \U_1\left(\fr{1}{\sqrt{2}}(t-\xb)\right)\otimes \U_2\left(\fr{1}{\sqrt{2}}(t+\xb)\right), 
\label{unitary-representation}
\eeq
whose spectrum is easily seen to coincide with $V_+$ as a consequence of property~4 from Definition~\ref{circle-theory}.  
We mention for future reference that if $\al_{(t,\xb)}(\,\cdot\,):=U(t,\xb)\,\,\cdot\,\, U(t,\xb)^*$ is the corresponding group of translation 
automorphisms and $\alb^{(1/2)}_{\t}(\,\cdot\, ):=\U_{1/2}(\t)\,\,\cdot\,\,\U_{1/2}(\t)^*$, then 
\beq
\alpha_{(t,\xb)}(A_1\otimes A_2) = \alb^{(1)}_{\fr{1}{\sqrt{2}}(t-\xb)}(A_1)\otimes \alb^{(2)}_{\fr{1}{\sqrt{2}}(t+\xb)}(A_2), 
\quad A_1\in \A_1,\, A_2\in\A_2. \label{automorphisms}
\eeq
Any double cone  $D\subset\RR^2$ can be expressed as a product of intervals
 on lightlines $D = \I_1\times \I_2$. We define the corresponding local von Neumann algebra by $\BvN(D):= \A_1(\I_1)\otimes\A_2(\I_2)$,
and for a general open region $\mco$ we put $\BvN(\mco)=\bigvee_{D\subset \mco}\BvN(D)$. The net of von Neumann algebras $(\BvN,U)$, 
which we call the chiral net,
satisfies all the properties from Definition~\ref{two-dim-net} except for the regularity property~5. Therefore, we introduce the following
weakly dense subnet of $C^*$-algebras
\beq
\mco\to\B(\mco):=\{\, A\in\BvN(\mco)\, |\, \lim_{\x\to 0}\|\al_{\x}(A)-A\|=0\, \}, 
\eeq
and denote the corresponding quasilocal algebra by $\B$. Then $(\B, U)$ is a local net of $C^*$-algebras in the sense of
Definition~\ref{two-dim-net}. We will call it the regular chiral net and refer to $(\A_1,\U_1)$, $(\A_2,\U_2)$  as its chiral 
components. We note for future reference that if $\B$ acts irreducibly on $\hil$, then $U$ is automatically the canonical representation of translations of this net (cf. Subsection~\ref{first-preliminaries}). Another useful fact is the obvious inclusion  
\beq
\hA_1\otimes_{\te{alg}}\hA_2\subset \B,
\eeq 
where $\otimes_{\te{alg}}$ is the algebraic tensor product.

Let $(\BvN,U)$ be a  chiral net, whose chiral components are $(\A_1,\U_1)$ and $(\A_2,\U_2)$.
Let $\pi_1$, $\pi_2$ be locally normal, covariant, positive energy representations of the respective
nets on $\real$. Then the  chiral net of $(\pi_1(\A_1), \U_{\pi_1})$, $(\pi_2(\A_2),\U_{\pi_2})$
is a covariant, positive energy representation of $(\BvN,U)$, which will be denoted by $(\pi(\BvN),U_{\pi})$, $\pi=\pi_1\otimes\pi_2$ and $\pi$ is called the product
representation of $\pi_1$ and $\pi_2$.
We note that $(\pi(\B),U_{\pi})$ is contained in the regular subnet of $(\pi(\BvN),U_{\pi})$. For faithful $\pi$ these
two nets coincide, due to Proposition~2.3.3 (2) of \cite{BR}. It is easily seen that $\pi$ is faithful (resp. irreducible),
if $\pi_1$ and $\pi_2$ are faithful (resp. irreducible). (Cf. Theorems 5.2 and 5.9 from Chapter IV of \cite{Ta}).

\subsection{Vacuum representations and asymptotic completeness}\label{scattering-two}

A regular chiral net $(\B,U)$ is in a vacuum representation, with the vacuum vector $\Om\in\hil$, if and
only if its chiral components  $(\A_1,\U_1)$, $(\A_2,\U_2)$ are in vacuum representations with
the respective vacuum vectors $\Om_1\in\K_1$, $\Om_2\in\K_2$ s.t. $\Om=\Om_1\otimes\Om_2$. (Cf. Proposition~\ref{invariant-vectors} below). In this subsection we show that any such regular chiral net has a complete particle interpretation 
in  terms of non-interacting Wigner particles. These facts follow from our results in \cite{DT10}, but
the argument below is more direct.

We start from the observation that the asymptotic fields have a particularly simple form
in chiral theories:
\bep\label{chiral-asymptotics} Let $(\A_1,\U_1)$, $(\A_2,\U_2)$ be two local nets of von Neumann algebras
in vacuum representations, with the respective vacuum vectors $\Om_1$, $\Om_2$.
Then, for any $A_1\in \hA_1$, $A_2\in\hA_2$ 
\beqa
\Phi_+^{\tout/\tin}(A_1\otimes A_2)&=&A_1\otimes (\Om_2|A_2\Om_2)I,\\
\Phi_-^{\tout/\tin}(A_1\otimes A_2)&=&(\Om_1|A_1\Om_1)I\otimes A_2.
\eeqa
\eep
\proof We consider only $\Phi_+^{\tout}$, as the remaining cases are analogous. From 
the defining relation~(\ref{asymptotic-field}) and formula~(\ref{automorphisms}),  we obtain
\beq
\Phi_+^{\tout}(A_1\otimes A_2)=\underset{T\to\infty}{\slim}\; A_1\otimes \int dt\, h_T(t)\alb_{\sqrt{2}t}^{(2)}(A_2).
\eeq
We set $A_2(h_T):=\int dt\, h_T(t)\alb_{\sqrt{2}t}^{(2)}(A_2)$. This sequence has the following properties:
\beqa
\lim_{T\to\infty}A_2(h_T)\Om_2&=&(\Om_2|A_2\Om_2)\Om_2, \label{auxiliary-one}\\
\lim_{T\to\infty}\|[A_2(h_T), A]\|&=&0, \textrm{ for any } A\in\hA_2.\label{auxiliary-two}
\eeqa
The first identity above follows from the mean ergodic theorem and the fact that $\Om_2$ is the only vector
invariant under the action of $\U_2$. The second equality is a consequence of the locality assumption from
Definition~\ref{circle-theory}. Since $\hA_2$ acts irreducibly, any  $\Psi\in\hilk_2$ has the 
form $\Psi=A\Om_2$ for some $A\in\hA_2$ \cite{Sa}. Thus we obtain from (\ref{auxiliary-one}), (\ref{auxiliary-two})
\beq
\underset{T\to\infty}{\slim}\; A_2(h_T)=(\Om_2|A_2\Om_2)I,
\eeq
which completes the proof. \qed\\
Now we can easily prove the main result of this subsection:
\bet\label{asymptotic-completeness-chiral} Any regular chiral net $(\B,U)$ in a vacuum representation is asymptotically complete. More precisely:
\beqa
& &\hil_+=\hilk_1\otimes \CC\Om_2, \label{hil-}\\
& &\hil_-=\CC\Om_1\otimes \hilk_2, \label{hil+}\\
& &\hil_+\pout\hil_-=\hil_+\pin\hil_-=\hil. \label{asymptotic-completeness-two}
\eeqa
Moreover, any such theory is non-interacting.
\eet
\begin{remark}
This result and Theorem~\ref{convergence-of-asymptotic-functionals} imply the convergence
of the asymptotic functional approximants $\{\si_{\Psi}^{(T)}\}_{T\in\real_+}$ for all $\Psi\in\hil$ of bounded energy in 
any regular chiral net in a vacuum representation.
\end{remark}
\proof  Using formula~(\ref{ergodic-theorem}) and the cyclicity of the vacuum $\Om$ under the action of $\B$, we obtain
\beq
\hil_{\pm}=[\,\Phi_{\pm}^{\tout}(F)\Om\,|\, F\in \B\,],
\eeq
where $[\,\cdot\,]$ denotes the norm closure.
Applying Proposition~\ref{chiral-asymptotics} and exploiting the cyclicity of $\Om_{1/2}$ under
the action of $\hA_{1/2}$, we obtain (\ref{hil-}) and (\ref{hil+}). The asymptotic completeness relation~(\ref{asymptotic-completeness-two}) 
also follows from Proposition~\ref{chiral-asymptotics}: For any $A_1\in\hA_1$, $A_2\in\hA_2$ 
\beqa
\Phi_{+}^{\tout}(A_1\otimes I)\Phi_{-}^{\tout}(I\otimes A_2)\Om=\Phi_{+}^{\tin}(A_1\otimes I)\Phi_{-}^{\tin}(I\otimes A_2)\Om
=A_1\Om_1\otimes A_2\Om_2.
\eeqa
Exploiting once again cyclicity of $\Om_{1/2}$, we obtain that scattering states are dense in the Hilbert space.

Now let us show the lack of interaction: Let $\Psi_{\pm}\in \hil_{\pm}$. Then, by (\ref{hil-}), (\ref{hil+}) and the
irreducibility of the action of $\hA_{1/2}$ on $\hilk_{1/2}$, there exist  $A_1\in\hA_1$, $A_2\in\hA_2$ s.t.
$\Psi_+=A_1\Om_1\otimes \Om_2$ and $\Psi_-=\Om_1\otimes A_2\Om_2$. Then
\beqa
\Psi_+\pout\Psi_-=\Phi_{+}^{\tout}(A_1\otimes I)\Phi_{-}^{\tout}(I\otimes A_2)\Om=A_1\Om_1\otimes A_2\Om_2\non\\
=\Phi_{+}^{\tin}(A_1\otimes I)\Phi_{-}^{\tin}(I\otimes A_2)\Om=\Psi_+\pin\Psi_-.
\eeqa
Hence the scattering operator, defined in (\ref{scattering-matrix}), equals the identity on $\hil$. \qed

\subsection{Charged representations and infraparticles}\label{infraparticles}

It is the goal of this subsection to clarify the particle content of chiral conformal field
theories in charged representations. More detailed particle properties of such theories, e.g.
superselection of direction of motion, will be studied in the next subsection.  

Let us first note the following simple relation between the single-particle subspaces
of a regular chiral net and the invariant vectors of its chiral components.
\bep\label{invariant-vectors} Let $(\A_1,\U_1)$, $(\A_2,\U_2)$ be local nets of von Neumann
algebras on $\real$. Then $\U_1$ (resp. $\U_2$)  has a non-trivial invariant vector, if and only if the 
single-particle subspace $\hil_-$ (resp. $\hil_+$) of the corresponding regular chiral 
net $(\B,U)$ is non-trivial.
\eep
\proof Suppose there exists a non-zero $\Om_{1}\in\hilk_1$, invariant under the action of $\U_1$. Then, for any $\Psi_{2}\in\hilk_{2}$, 
\beqa
U(t,-t)(\Om_{1}\otimes \Psi_2)&=&\Om_{1}\otimes \Psi_2,
\eeqa
for $t\in\real$. Hence the subspace $\hil_{-}$ is non-trivial. Similarly, the existence of a non-zero $\Om_2\in\hilk_2$, invariant under
the action of $\U_2$, implies the non-triviality of $\hil_+$.
 
Now suppose $\Psi\in\hil_{-}$ and  $\U_{1}$ has no non-trivial, invariant vectors.  
Then, by the mean ergodic theorem,
\beqa
\Psi= \lim_{T\to\infty}\fr{1}{T}\int_0^T dt\, U(t,-t)\Psi
=\lim_{T\to\infty}\fr{1}{T}\int_0^T dt\, \big(\U_{1}(\sqrt{2}t)\otimes I\big)\Psi=0.
\eeqa
Thus we established that $\hil_-=\{0\}$. Similarly, the absence of non-trivial, invariant vectors of $\U_2$ implies
that $\hil_+=\{0\}$. \qed\\
Let $(\B, U)$ be a regular chiral net in  a charged irreducible (product) representation.
That is $\B$ acts irreducibly on a non-trivial Hilbert space, which has the tensor product structure, by
our definition of chiral nets, and does not contain 
non-zero invariant vectors of $U$. The particle structure of such theories is described
by the following theorem.
\bet\label{existence-of-particles} Let $(\B, U)$ be a regular chiral net in a charged irreducible (product) representation
acting on a Hilbert space $\hil$. Then:
\begin{enumerate}
\item[(a)] $\hil_{+}=\{0\}$ or $\hil_-=\{0\}$ i.e. the theory does not describe Wigner particles.
\item[(b)] For any non-zero vector $\Psi\in P_{E}\hil$, $E\geq 0$, all the limit points of the 
net  $\{\si_{\Psi}^{(T)}\}_{T\in\real_+}$, given by (\ref{approximants}), are different from zero. 
\end{enumerate}
Hence $(\B,U)$ describes infraparticles.
\eet
\proof Part (a) follows immediately from Proposition~\ref{invariant-vectors} and the absence of non-zero invariant vectors of $U$ in $\hil$. 
As for part (b), since $\B$ acts irreducibly on $\hil$, its chiral components $(\A_{1/2} ,\U_{1/2})$  act irreducibly on
their respective Hilbert spaces $\K_{1/2}$.
We note that for any non-zero vector $\Psi\in P_E\hil$ we can find 
a sequence of vectors $\{\Psi_{n}\}_{n\in\nat}$  from $\K_1$ s.t.  $\Psi_1\neq 0$ and
\beq
(\Psi|(C\otimes I)\Psi)=\sum_{n\in\nat}(\Psi_n|C\Psi_n) \label{PsiC}
\eeq
for all $C\in B(\K_1)$. Moreover, we can assume without loss of generality that $\K_1$ does not contain non-trivial invariant vectors  
of $\U_1$. Then we obtain from Lemma~\ref{constant1}~(b) the existence of a local
operator $A\in \A_1$ and $f\in S(\real)$ s.t. $\supp\,\tf\cap\real_+=\emptyset$, which satisfy $A(f)\Psi_1\neq 0$. 
We note that any $B:=A(f)\otimes I$ is a non-zero element of $\B$ which is almost local and energy decreasing. 
Consequently, $B^*B$ belongs to the algebra of detectors $\mcC$  of the net $(\B,U)$. We consider the corresponding asymptotic functional
approximants
\beqa
\si_{\Psi}^{(T)}(B^*B)&=&\int dt\, h_T(t)\int d\xb\, (\Psi| \al_{(t,\xb)}(B^*B)\Psi)\non\\
&=&\int dt\, h_T(t)\int d\xb\,(\Psi|\alb_{ (\sqrt{2})^{-1}(t-\xb)}^{(1)}(A(f)^*A(f))\otimes I)\Psi )\non\\
&\geq &\int d\xb\,(\Psi_1|(\alb^{(1)}_{(\sqrt{2})^{-1}\xb} (A(f)^*A(f)))\Psi_1)\neq 0,
\eeqa
where in the last step we made use of (\ref{PsiC}). As the last expression is independent of $T$, all the limit points
of $\{ \si_{\Psi}^{(T)}\}_{T\in\real_+}$ are different from zero. \qed

\subsection{Infraparticles with superselected direction of motion} \label{infra}

In Theorem~\ref{existence-of-particles} above we have shown that any
charged irreducible (product) representation of a chiral conformal field theory contains infraparticles. 
In this subsection we demonstrate that in a large class of examples these infraparticles have superselected 
direction of motion in the sense of Definition~\ref{superselection}.

Let $(\A,\U)$ be a local net of von Neumann algebras on $\real$, acting on a  Hilbert space $\hilk$. We assume that
this net is in a vacuum representation, with the vacuum vector $\Omz\in\hilk$. Let $W$ be a unitary operator on $\hilk$ which 
implements a symmetry of this net i.e.
\beqa
& &W\A(\I)W^{*}\subset \A(\I), \label{symmetry-one} \\
& &W\U(t)W^*= \U(t), \label{symmetry-two} \\
& &W\Om_0=\Om_0,  \label{symmetry-three}
\eeqa
for any open, bounded  interval $\I\subset \real$ and any $t\in\real$. We assume that $W$ gives rise to
a non-trivial representation of the  group $\mathbb{Z}_2$ i.e. $\Ad W\neq \id$ and $W^2=I$.
We define the subspaces
\beqa
\A_{\ev}(\I)&=&\{\, A\in\A(\I)\,|\, WAW^*=A\,\},\\
\A_{\odd}(\I)&=&\{\, A\in\A(\I)\,|\, WAW^*=-A\,\}.
\eeqa 
Let $\A_{\ev}$ (resp. $\A_{\odd}$) be the norm-closed linear span of all operators from some $\A_{\ev}(\I)$ (resp. $\A_{\odd}(\I)$), 
$\I\subset\RR$. Clearly, $(\A_{\ev}, \U)$ is again a local net of von Neumann algebras on the real line acting on $\hilk$. 
We introduce the subspaces $\hilk_{\ev}=[\A_{\ev}\Om_0]$, $\hilk_{\odd}=[\A_{\odd}\Om_0]$, where $[\,\cdot\,]$ denotes
the closure, which are invariant under
the action of $\A_{\ev}$ and $\U$, and satisfy $\hilk=\hilk_{\ev}\oplus\hilk_{\odd}$. $\hilk_{\odd}$ gives rise to  the representation
\beqa
\pi_{\odd}(A) &=& A|_{\hilk_{\odd}},\quad A\in\A_{\ev},\label{odd-def}\\
\U_{\odd}(t)&=&\U(t)|_{\hilk_{\odd}},\quad t\in\real.
\eeqa
Its relevant properties are summarized in the following lemma, which we  prove in Appendix~\ref{Appendix-B}. 
\bel\label{odd-lemma} $(\pi_{\odd},\hilk_{\odd})$ is a covariant, positive energy representation of
$(\A_{\ev},\U)$, in which the translation automorphisms
are implemented by $\U_{\odd}$. Moreover:
\begin{enumerate}
\item[(a)]  $\pi_{\odd}$ is a locally normal, faithful and irreducible representation of $\A_{\ev}$.
\item[(b)]   $\U_{\odd}$ does not admit non-trivial invariant vectors. 
\end{enumerate}
\eel
\noindent We set $\At:=\pi_{\odd}(\A_{\ev})$, $\Ut(t):=\U_{\odd}(t)$. By the above lemma $(\At,\Ut)$ is again a local
net of von Neumann algebras on the real line. We  define its representation on $\hilk_{\ev}$
\beqa
\pi_{\ev}(\Att) &=& \pi_{\odd}^{-1}(\Att)|_{\hilk_{\ev}},\quad \Att\in\At,\label{even-def}\\
\U_{\ev}(t)&=&\U(t)|_{\hilk_{\ev}},\quad t\in\real \label{even-def-U}
\eeqa
and state the following fact, whose proof is given in Appendix~\ref{Appendix-B}.
\bel\label{even-lemma} $(\pi_{\ev},\hilk_{\ev})$ is a covariant, positive energy representation of
 $(\At,\Ut)$, in which the translation automorphisms
are implemented by $\U_{\ev}$. Moreover
\begin{enumerate}
\item[(a)]  $\pi_{\ev}$ is a locally normal, faithful and irreducible representation of $\At$.
\item[(b)]   $\U_{\ev}$  admits a unique (up to a phase) invariant vector, which is cyclic for any $\pi_{\ev}(\At(\I))$. 
\end{enumerate}
\eel
\noindent We conclude that $(\pi_{\ev}(\At),\U_{\ev})$ is a local net of von Neumann algebras in a vacuum representation
with the vacuum vector $\Omz\in\hilk_{\ev}$.

We remark that the above abstract construction can be performed in a number of concrete cases. 
If a M\"obius covariant net $\I\to\A(\I)$ on $S^1$, in a vacuum representation, admits an 
automorphism\footnote{An automorphism $\gamma$ of a net $\A$ is an automorphism of the quasilocal algebra $\A$ which
preserves each local algebra $\A(\I)$.} $\gamma$
of order 2 which preserves the vacuum state, then one can define $W$ by
\beq
WA\Omz = \gamma(A)\Omz,\quad  A \in \A(\I).
\eeq
This does not depend on the choice of the interval $\I$ and defines a unitary operator
thanks to the invariance of the vacuum state. This $W$ automatically commutes with
the action of the M\"obius group (in particular with the action of translations) 
as a consequence of the Bisognano-Wichmann property \cite{GF93}. Thus, upon restriction 
to the real line, we obtain a local net equipped with a unitary $W$ which satisfies~(\ref{symmetry-one})-(\ref{symmetry-three}).    
Non-trivial automorphisms $\ga$ appear, in particular, in the $U(1)$-current net ($\gamma: J(z) \to -J(z)$) \cite{BMT88},
in loop group nets of a compact group $G$ with a $\ZZ_2$-subgroup in $G$ \cite{Wa98} and
in the tensor product net $\A\otimes \A$ for an arbitrary M\"obius covariant net $\A$, where $\ga$ is the flip symmetry.

Coming back to the abstract setting, we introduce the class of two-dimensional theories, we are interested in: 
Let $(\A_1,\U_1)$, $(\A_2,\U_2)$
be two local nets of von Neumann algebras on $\real$, in vacuum representations, acting on Hilbert spaces $\hilk_1$, $\hilk_2$. 
We denote the respective vacuum vectors by $\Om_1$, $\Om_2$ and introduce the corresponding regular chiral net $(\B,U)$.
We assume the existence of unitaries $W_1$, $W_2$, which give rise to non-trivial representations
of $\mathbb{Z}_2$  and implement symmetries of the respective nets on $\real$ as defined in 
(\ref{symmetry-one})-(\ref{symmetry-three}). By the construction described above we obtain the nets  
$(\At_1,\Ut_1)$, $(\At_2,\Ut_2)$, acting on $\hilk_{1,\odd}$, $\hilk_{2,\odd}$. We denote by $(\BttvN, \Utt)$  the corresponding 
chiral net acting on $\hiltt=\hilk_{1,\odd}\otimes\hilk_{2,\odd}$ and by $(\Btt,\Utt)$ its regular subnet. Let us summarize
its properties.
\bep\label{asymptotic-representations} The regular chiral net $(\Btt,\Utt)$, whose chiral components are 
$(\At_1,\Ut_1)$, $(\At_2,\Ut_2)$, has the following properties:
\begin{enumerate}
\item[(a)] $\Btt$ acts irreducibly on $\hiltt$. 

\item[(b)] $(\Btt,\Utt)$ does not admit Wigner particles ($\hiltt_{\pm}=\{0\}$), but all the asymptotic
functionals of the form $\{\, \psi_\Psi^{\tout} \,|\,\Psi\neq 0,\,\Psi\in P_E\hiltt,\, E\geq 0 \,\}$ are non-zero.

\item[(c)] $\pi_{\r}=\iota_1\otimes\pi_{2,\ev}$ and $\pi_{\l}=\pi_{1,\ev}\otimes\iota_2$ are irreducible, faithful, covariant representations of $(\Btt,\Utt)$, acting on $\hil_{\pir}:=\hilk_{1,\odd}\otimes\hilk_{2,\ev}$ and 
$\hil_{\pil}:=\hilk_{1,\ev}\otimes\hilk_{2,\odd}$, respectively. The respective (canonical) unitary representations of translations are given by $U_{\pi_\r}(x):=U(x)|_{\hil_{\pir}}$ and $U_{\pi_\l}(x):=U(x)|_{\hil_{\pil}}$.

\item[(d)] $\hil_{\pir,-}=\{0\}$ and $\hil_{\pir,+}\neq \{0\}$ while $\hil_{\pil,-}\neq \{0\}$ and $\hil_{\pil,+}=\{0\}$. Consequently,
$\pi_{\r}$ is not unitarily equivalent to $\pi_{\l}$.

\end{enumerate}
In part (c) $\iota_{1/2}$ are the defining representations of $\At_{1/2}$. Representations $\pi_{1/2,\ev}$ are defined as 
in (\ref{even-def}), (\ref{even-def-U}). 
\eep
\proof  Part (a) follows from the irreducibility of $\pi_{1/2,\odd}$, shown in Lemma~\ref{odd-lemma}.
As for part~(b), we obtain from Lemma~\ref{odd-lemma}~(b) and Proposition~\ref{invariant-vectors} that $\hiltt_{\pm}=\{0\}$.
On the other hand, Theorem~\ref{existence-of-particles} ensures that the relevant asymptotic functionals are non-zero.
Irreducibility and faithfulness of $\pi_{\r/\l}$ in part (c) follow from  Lemma~\ref{even-lemma}~(a) and Lemma~\ref{odd-lemma}~(a).
Proceeding to part (d), we note that, by faithfulness of $\pi_{\r}$,
the net  $(\pi_{\r}(\Btt), U_{\pi_{\r}})$ coincides with the regular chiral subnet of $(\pi_{\r}(\BttvN), U_{\pi_{\r}})$, whose chiral components are $(\At_1,\Ut_1)$ and $(\pi_{2,\ev}(\At_2),\U_{2,\ev})$. From Lemma~\ref{odd-lemma}~(b),
Lemma~\ref{even-lemma}~(b) and Proposition~\ref{invariant-vectors} we obtain that $\hil_{\pir,-}=\{0\}$ and $\hil_{\pir,+}\neq \{0\}$. An
analogous reasoning, applied to $\pi_{\l}$, shows that $\hil_{\pil,-}\neq \{0\}$ and $\hil_{\pil,+}=\{0\}$. Hence, due to
relation~(\ref{equivalence-two}), the two nets are not unitarily equivalent. \qed\\
In view of part~(b) of the above proposition, the theory $(\Btt,\Utt)$ describes infraparticles. In the following
theorem, which is our main result, we show that these infraparticles have superselected direction of motion, in the
sense of Definition~\ref{superselection}.
\bet\label{representation-decomposition-two} Consider the regular chiral net $(\Btt,\Utt)$,  
constructed above. Let $\psi\in \{\, \psi_\Psi^{\tout} \,|\,\Psi\neq 0,\,\Psi\in P_E\hiltt,\, E\geq 0 \,\}$ and
let $\pi_{\psi}$ be its GNS representation.  
Then $\pi_{\psi}$ is a type~I representation with atomic center. It
contains both right-moving and left-moving irreducible subrepresentations which are unitarily
equivalent to $\pi_{\r}$ and $\pi_{\l}$, respectively. 
Hence the theory describes infraparticles with 
superselected  direction of motion.
\eet
\begin{remark} Let us consider the regular chiral net $(\B,U)$  in the vacuum representation. Then, similarly as in the theorem above,
the GNS representation $\pi_{\psi}$ induced by any particle weight 
$\psi\in  \{\, \psi_\Psi^{\tout} \,|\, \Psi\notin\CC\Om, \Psi\in P_E\hil, E\geq 0 \,\}$
is of type~I with atomic center.  However,
any non-trivial irreducible subrepresentation of $\pi_{\psi}$ is unitarily equivalent to the
defining vacuum representation i.e. $\psi$ is  neutral. These facts
are easily verified by modifying the proof below.    
\end{remark}
\proof Let us  first consider the regular chiral  net $(\B,U)$ acting on $\hil$. By Theorem~\ref{asymptotic-completeness-chiral},  
 $\K_+:=\K_{1,\odd}\otimes \CC\Om_2\subset \hil_{+}$ and $\K_-:=\CC \Om_1\otimes \K_{2,\odd}\subset\hil_{-}$.
Any vector  $\Psi\in P_E(\K_+\pout \K_-)$, $E\geq 0$, gives rise to functionals $\rho_{\Psi,\pm}$, defined by (\ref{density+}), 
(\ref{density-}). They have the form
\beq
\rho_{\pm,\Psi}(\,\cdot\,)=\sum_{n\in\nat} (\Psi_{\pm,n}|\,\cdot\, \Psi_{\pm,n}), 
\eeq
where $\Psi_{\pm,n}\in P_E \K_{\pm}$. (Cf. formula~(\ref{reduced-matrix}) and the subsequent discussion). Since $\Psi\neq 0$,
we can assume that $\Psi_{+,1}\neq 0$ and $\Psi_{-,1}\neq 0$.
We also note for future reference that $\K_+\subset\hil_{\pir}$ and $\K_-\subset\hil_{\pil}$.

Let us now proceed to the net $(\Btt,\Utt)$, acting on $\hiltt=\K_+\pout \K_-\subset \hil$, and let $\mcLtt$ be
the left ideal of $\Btt$, given by definition~(\ref{left-ideal}). For any $\Ltt\in\mcLtt$ we define 
\beq
L=(\pi_{1,\odd}^{-1}\otimes\pi_{2,\odd}^{-1})(\Ltt)\in\mcL, \label{LvsLtt} 
\eeq
where $\mcL$ is the corresponding left ideal of $\B$. We note that such $L$ leaves
the subspaces $\hil_{\pir}$ and $\hil_{\pil}$ invariant.
Exploiting Theorem~\ref{convergence-of-asymptotic-functionals} and 
formula~(\ref{reduced-matrix}), we obtain
\beqa
\psi_{\Psi}^{\tout}(\Ltt_1,\Ltt_2)=\sum_{n\in \nat} \int d\xb\, \{ (\Psi_{+,n}|(L_1^*L_2)(\xb)\Psi_{+,n})
+(\Psi_{-,n}|(L_1^*L_2)(\xb)\Psi_{-,n}) \}. \label{main-technical-two}
\eeqa
It follows from Theorem~\ref{integrability} that for any $L$ given by (\ref{LvsLtt}) the Fourier transforms
\beqa
L\ti{\Psi}_{+,n}(\pb)&:=&(2\pi)^{-1/2} \int d\xb\, e^{-i\pb\xb} LU_{\pi_\r}(\xb)^*\Psi_{+,n}, \label{rightFT} \\ 
L\ti{\Psi}_{-,n}(\pb)&:=&(2\pi)^{-1/2} \int d\xb\, e^{i\pb\xb} LU_{\pi_\l}(\xb)^*\Psi_{-,n} \label{leftFT}
\eeqa
belong to $\hil_{\pir}\otimes L^2(\real_+,d\pb)$ and $\hil_{\pil}\otimes L^2(\real_+,d\pb)$ respectively.
Since $\pi_{\r}(\Btt)$ acts irreducibly on $\hil_{\pir}$ and $U_{\pi_\r}$ does not have non-zero invariant vectors,
we obtain from Lemma~\ref{constant1}~(a) the existence of $\Ltt_+\in\mcLtt$ s.t.
$L_+\Psi_{+,1} \neq 0$. Since $L_+ U_{\pi_\r}(\xb)^*\Psi_{+,1}$ is a continuous
function of $\xb$, it is nonzero as a square-integrable function, hence
$\{L_+\ti{\Psi}_{+,1}(\pb)\}_{\pb\in\real_+}\neq 0$.
Analogously, we can find $\Ltt_-\in\mcLtt$ s.t. $\{L_-\ti{\Psi}_{-,1}(\pb)\}_{\pb\in\real_+}\neq 0$.
For future reference, we note the equalities
\beqa
\al_{x}(L)\ti{\Psi}_{+,n}(\pb)&=& e^{-i(\pb,\pb)x}U_{\pi_\r}(x)L\ti{\Psi}_{+,n}(\pb), \label{rightFTp} \\ 
\al_{x}(L)\ti{\Psi}_{-,n}(\pb)&=& e^{-i(\pb,-\pb)x}U_{\pi_\l}(x)L\ti{\Psi}_{-,n}(\pb),  \label{leftFTp}
\eeqa
which hold in the sense of $\hil_{\pir}\otimes L^2(\real_+,d\pb)$ and
$\hil_{\pil}\otimes L^2(\real_+,d\pb)$ respectively.
These relations are easily verified for such $\Psi_{\pm,n}\in P_{E}\K_{\pm}$
that $\real\ni \xb \to LU_{\pi_{\r/\l }}(\xb)^*\Psi_{\pm,n}$ decay rapidly in norm as $|\xb|\to \infty$, since in this case the Fourier transform is pointwise defined.
The general case
follows from the fact that such vectors form a dense subspace in $P_{E}\K_{\pm}$ (cf. formula~(\ref{rapid-decay}) below) 
and that the maps $P_{E}\K_{\pm}\ni \Psi\to \{ L\ti{\Psi}(\pb)\}_{\pb\in\real_+}\in\hil_{\pi_{\r/\l}}\otimes L^2(\real_+,d\pb)$
are norm-continuous. This latter fact is a consequence of Theorem~\ref{integrability} and the (Hilbert space valued) Plancherel theorem.

After this preparation we study the structure of the GNS representation induced by $\psi$.
Let us first consider the following auxiliary representation of $(\Btt,\Utt)$ 
\beq
\pi_{1}(\,\cdot\,):=\bigoplus_{n\in\nat}\big( \{\pi_{\r}(\,\cdot\,)\otimes I\}\oplus \{\pi_{\l}(\,\cdot\,)\otimes I\}\big),\label{auxiliary-rep}
\eeq
acting on $\hil_{\pi_1}:=\bigoplus_{n\in\nat}\big( \{\hil_{\pir}\otimes L^2(\real_+,d\pb) \}\oplus \{\hil_{\pil}\otimes L^2(\real_+,d\pb)\} \big)$. 
 From definition~(\ref{auxiliary-rep}) and relation~(\ref{double-commutant}) we conclude that $\pi_1$ and its subrepresentations are 
of type~I with atomic center. Moreover, $\pi_1$ is covariant and it is easily seen that the canonical representation of 
translations is given by
\beq
U^{\can}_{\pi_1}(x)=\bigoplus_{n\in\nat}\big( \big\{U_{\pi_\r}(x)\otimes I\big\}\oplus \big\{U_{\pi_\l}(x)\otimes I\big\}).
\eeq
We note that $\pi_{\psi}$ is unitarily equivalent to a subrepresentation of $\pi_1$.
In fact, the map $W_1:\hil_{\pi_\psi}\to \hil_{\pi_1}$, given by
\beq
W_1: |\Ltt\> \to \bigoplus_{n\in\nat } \big(  \{\, L\ti\Psi_{+,n}(\pb)\,\}_{\pb\in\real_+} \oplus
   \{\, L\ti\Psi_{-,n}(\pb) \,\}_{\pb\in\real_+}  \big), \label{W1def}
\eeq
intertwines the two representations and is an isometry by formula~(\ref{main-technical-two})
and the (Hilbert space valued) Plancherel theorem.  
It is easily checked that the canonical representation of translations $U_{\pi_{\psi}}^{\can}$ in the
representation $\pi_{\psi}$ is given by the relation
\beq
W_1 U^{\can}_{\pi_{\psi}}(x)=U^{\can}_{\pi_1}(x)W_1. 
\eeq
Recalling that $\V_{\pi_{\psi}}(x)=U^{\can}_{\pi_{\psi}}(x)U_{\pi_{\psi}}^{-1}(x)$, where $U_{\pi_{\psi}}$ is given by~(\ref{standard}),
we obtain
\beqa
W_1 \V_{\pi_{\psi}}(x)|\Ltt\> 
&=&W_1 U^{\can}_{\pi_{\psi}}(x)|\al_{-x}(\Ltt)\> 
=U^{\can}_{\pi_1}(x) W_1|\al_{-x}(\Ltt)\> \non\\
&=& \bigoplus_{n\in \nat }\big( \{\, U_{\pi_\r}(x)\al_{-x}(L)  \ti\Psi_{+,n}(\pb) \,\}_{\pb\in\real_+}\oplus
\{\,  U_{\pi_\l}(x) \al_{-x}(L) \ti\Psi_{-,n}(\pb) \,\}_{\pb\in\real_+} \big) \non\\   
&=& \bigoplus_{n\in\nat}\big( \{\,I\otimes e^{i(\pb, \pb)\x} \}_{\pb\in\real_+}\oplus 
\{ I\otimes e^{i(\pb,-\pb)\x} \}_{\pb\in\real_+} \big) W_1|\Ltt\>, \label{V-relation}
\eeqa
where in the last step we made use of relations~(\ref{rightFTp}), (\ref{leftFTp}). Now let $\bQ$  be the 
generator of space translations of $\V_{\pi_{\psi}}$ and let $\hil_{\pi_\psi,\rr}$ (resp. $\hil_{\pi_\psi,\ll}$) be its spectral subspace
corresponding to the interval $[0,\infty)$ (resp. $(-\infty,0)$). Then, by formula~(\ref{V-relation}),
\beqa
& &W_1\hil_{\pi_\psi,\rr}=P_{\r} W_1\hil_{\pi_\psi},\\ 
& &W_1\hil_{\pi_\psi,\ll}=P_{\l} W_1\hil_{\pi_\psi}, 
\eeqa
where $P_{\r/\l}$ are the projections on the subspaces $\bigoplus_{n\in\nat} \{\hil_{\pi_{\r/\l}}\otimes L^2(\real_+,d\pb) \}$ in 
$\hil_{\pi_1}$. From definition~(\ref{W1def}) and the remarks after formula~(\ref{leftFT}) we conclude that 
$P_{\r} W_1\hil_{\pi_\psi}\neq \{0\}$ and $P_{\l} W_1\hil_{\pi_\psi}\neq \{0\}$. Consequently $\pi_{\psi}$ has both right-moving and left-moving irreducible subrepresentations.

Let $\pi$ be an irreducible subrepresentation of $\pi_{\psi}$, acting on a non-trivial subspace $\K\subset\hil_{\pi_\psi,\rr}$ (i.e. a right-moving subrepresentation). Then $W_1\pi(\,\cdot\,)W_1^*$ is an irreducible subrepresentation of $\pi_\r(\,\cdot\,)\otimes I$
acting on $\hil_{\pir}\otimes \big( \bigoplus_{n\in\nat}L^2(\real_+,d\pb)\big)$. By irreducibility, we conclude that $\pi$ 
is unitarily equivalent to  $(\pi_{\r}(\,\cdot\,)\otimes I)|_{\hil_{\pir}\otimes \CC e}$ for some non-zero $e\in \bigoplus_{n\in\nat}L^2(\real_+,d\pb)$. This latter representation can be identified with $\pi_{\r}$. An analogous argument shows that any left-moving irreducible 
subrepresentation of $\pi_{\psi}$ is unitarily equivalent to $\pi_{\l}$. Hence, by Proposition~\ref{asymptotic-representations}~(d),
$(\Btt,\Utt)$ describes infraparticles with superselected direction of motion.  \qed\\ 
Let us assume for a moment that $\hiltt$ is separable. Then we obtain from the above theorem and  formula~(\ref{direct-int}) that the GNS representation of any particle weight $\psi^{\tout}_{\Psi}$  of the net $(\Btt,\Utt)$, where $\Psi\neq 0$ is a vector of bounded energy, has the form
\beqa
\pi_{\psi^{\tout}_{\Psi}}\simeq \int^{\oplus}_{Z_{\r}}d\mu_{\r}(z)\,\pi_{\r,z}  
\oplus \int^{\oplus}_{Z_{\l}}d\mu_{\l}(z)\,\pi_{\l,z}.
\eeqa
Here $(Z_{\r/\l}, d\mu_{\r/\l} )$ are some Borel spaces and $\pi_{\r/ \l,z}=\pi_{\r/\l}$ for all $z\in Z_{\r/\l}$.
A decomposition of $\psi^{\tout}_{\Psi}$ into pure particle weights, which induce the irreducible representations appearing in the decomposition of $\pi_{\psi^{\tout}_{\Psi}}$, was obtained by Porrmann in \cite{Po04.1,Po04.2} 
(cf. formula~(\ref{decomposition-first}) above). However, to apply Porrmann's abstract argument, one has to restrict attention to 
countable (resp. separable) subsets of all the relevant objects and it is not guaranteed that the resulting (restricted) pure particle weights  extend to the original domains. It is therefore worth pointing out that the theory $(\Btt,\Utt)$ admits a large class of particle weights, whose decomposition can be performed in the original framework. To our knowledge this is the first such decomposition in the presence of infraparticles. (See however \cite{Jo91} for some partial results on the Schroer model). These particle weights belong to the set
$\{\,\psi^{\tout}_{\Psi}\,|\, \Psi\in\Dom\,\}$, where $\Dom\subset\hiltt$ is a dense domain spanned  
by vectors of the form 
\beq
\Psi=F_1\Om_1\otimes F_2\Om_2, \label{odd-vector}
\eeq
where $F_1\in \A_{1,\odd}, F_2\in\A_{2,\odd}$ are s.t. $F_1\otimes I, I\otimes F_2\in \B$ are almost local 
and have compact energy-momentum transfer (see formula~(\ref{energy-momentum-transfer})). 
The proof of the following proposition exploits some ideas from \cite{En75}.
\bep\label{infraparticle-decomposition} Consider the regular chiral net $(\Btt,\Utt)$
constructed above. Denote by $\mcLtt$ its left ideal, given by definition (\ref{left-ideal}).
Then, for any non-zero vector $\Psi\in\Dom$, there exist 
continuous fields of pure particle weights $\De_{\r,n}\ni \pb\to\psi_{\r,n,\pb}(\,\cdot\,,\,\cdot\,)$ and 
$\De_{\l,m}\ni \pb\to\psi_{\l,m,\pb}(\,\cdot\,,\,\cdot\,)$ s.t. for any $\Ltt_1,\Ltt_2\in \mcLtt$ 
\beqa
\psi_{\Psi}^{\tout}(\Ltt_1,\Ltt_2)=\sum_{n\in\Sub_{\r}}\int_{\De_{\r,n}} d\pb\, \psi_{\r,n,\pb}(\Ltt_1,\Ltt_2)
+ \sum_{m\in\Sub_{\l}}\int_{\De_{\l,m}} d\pb\, \psi_{\l,m,\pb}(\Ltt_1,\Ltt_2),
\eeqa
where $\Sub_{\r}, \Sub_{\l}\subset\nat$ are non-empty finite subsets and  $\De_{\r,n},\De_{\l,m}\subset \real_+$ are non-empty, open subsets  for any $n\in\Sub_{\r}$, $m\in\Sub_{\l}$. 
Moreover:
\begin{enumerate}
\item[(a)] The characteristic energy-momentum vectors of the weights $\psi_{\r,n,\pb}$ (resp. $\psi_{\l,m,\pb}$) 
are equal to $q_{\r,n,\pb}=(\pb, \pb)$ (resp. $q_{\l,m,\pb}=(\pb, -\pb)$).
\item[(b)] The GNS representation induced by any $\psi_{\r,n,\pb}$ (resp. $\psi_{\l,m,\pb}$) is unitarily equivalent to $\pi_{\r}$
(resp. $\pi_{\l}$).  
\end{enumerate}
The representations $\pi_{\r/\l}$ appeared in Proposition~\ref{asymptotic-representations}.
\eep   
\begin{remark} Parts (b) and (d) of Proposition~\ref{asymptotic-representations} show that 
spectral properties of the energy-momentum operators in the representations induced by the pure particle
weights $\psi_{\r/\l,n,\pb}$ are different from those in the original representation: In the case of $U_{\pi_{\r}}$ the right branch of
the lightcone contains the singularities characteristic for Wigner particles, while in the left branch
such singularities are absent. (For $U_{\pi_{\l}}$ the opposite situation occurs). For  infraparticles
in physical spacetime a more radical version of this phenomenon may occur: There one expects isolated singularities 
at the characteristic energy-momentum values of the respective pure particle weights.
(Cf. Section 2 (iii) of \cite{BPS91}).
\end{remark}
\proof Any vector  $\Psi\in\Dom$ has the form
\beq
\Psi=\sum_{k,l}c_{k,l} F_{\r,k}\Om_1\otimes F_{\l,l}\Om_2,
\eeq
where the sum is finite and $F_{\r,k}$, $F_{\l,l}$ have properties specified below formula~(\ref{odd-vector}). 
Applying the Gram-Schmidt procedure, we can ensure that the 
systems of vectors $\{F_{\r,k}\Om_1\}_{k=0}^{M}$, $\{F_{\l,l}\Om_2\}_{l=0}^N$ are orthonormal. 
Since $\hiltt=\hilk_{1,\odd}\otimes\hilk_{2,\odd}\subset \hilk_1\otimes\hilk_2=\hil$, we can write
\beq
\Psi=\sum_{k,l}c_{k,l}\Phip(F_{\r,k}\otimes I)\Phim(I\otimes F_{\l,l})\Om,
\eeq
where we made use of Proposition~\ref{chiral-asymptotics} applied to the net $(\B,U)$. For any $\Ltt\in\mcLtt$ we define $L=(\pi_{1,\odd}^{-1}\otimes\pi_{2,\odd}^{-1})(\Ltt)\in\mcL$, where $\mcL$ is the left ideal of $\B$, given by definition~(\ref{left-ideal}).
In view of Theorem~\ref{convergence-of-asymptotic-functionals}, we get
\beqa
\psi_{\Psi}^{\tout}(\Ltt_1,\Ltt_2)&=&\sum_{n\in\Sub_{\r}} \int d\xb\, ((\FF_{\r,n}\otimes I)\Om |(L_1^*L_2)(\xb) (\FF_{\r,n}\otimes I)\Om )\non\\
&+&\sum_{m\in\Sub_{\l}}\int d\xb\, ((I\otimes {\FF}_{\l,m})\Om |(L_1^*L_2)(\xb)(I\otimes {\FF}_{\l,m})\Om), \label{starting-point}
\eeqa
where $\FF_{\r,n}=\sum_k  c_{k,n}F_{\r,k}$, $\FF_{\l,m}=\sum_{l}c_{m,l}F_{\l,l}$ and the sets $\Sub_{\r}$ and $\Sub_{\l}$ are chosen
so that $\Psi_{\r,n}:=(\FF_{\r,n}\otimes I)\Om\neq 0$ and $\Psi_{\l,m}:=(I\otimes \FF_{\l,m})\Om\neq 0$ for $n\in \Sub_{\r}$ and 
$m\in\Sub_{\l}$. We note that both sets are non-empty, if $\Psi\neq 0$. (Cf. formula~(\ref{reduced-matrix}) and the subsequent remarks).

Let us consider the first sum in (\ref{starting-point}) above:
Any $L\in\mcL$ is a finite linear combination of operators of the form $AB$, where $A,B\in\B$ and $B$ is almost local and energy decreasing.
Since we assumed that $F_{\r,k}\otimes I$ are almost local, the functions
\beqa
\real\ni \xb\to ABU_{\pi_{\r}}(\xb)^*(\FF_{\r,n}\otimes I)\Om=A[B, (\FF_{\r,n}\otimes I)(-\xb)]\Om \label{rapid-decay}
\eeqa
decrease in norm faster than any inverse power of $|\xb|$. Consequently, the Fourier transform
\beqa
L\ti{\Psi}_{\r,n}(\pb):=(2\pi)^{-1/2} \int d\xb\, e^{-i\pb\xb} LU_{\pi_{\r}}(\xb)^*(\FF_{\r,n}\otimes I)\Om \label{Psi-definition}
\eeqa 
is a norm-continuous function. It is compactly supported in $\real_+$ due to the spectrum condition and the fact that  
the energy-momentum transfer of each $\FF_{\r,n}\otimes I$ is bounded.  
By the (Hilbert space valued) Plancherel theorem, we can write
\beq
\int d\xb\, ((\FF_{\r,n}\otimes I)\Om |(L_1^*L_2)(\xb)(\FF_{\r,n}\otimes I)\Om)=\int d\pb\, (L_1\ti{\Psi}_{\r,n}(\pb)|L_2\ti{\Psi}_{\r,n}(\pb)).
\eeq
We define
\beqa
\psi_{\r,n,\pb}(\Ltt_1,\Ltt_2):=(L_1\ti{\Psi}_{\r,n}(\pb)|L_2\ti{\Psi}_{\r,n}(\pb)).
\eeqa
It is easy to see that non-zero $\psi_{\r,n,\pb}$ are particle weights in the sense of Definition~\ref{weight-def}: Positivity and property~1
are obvious. The continuity requirement in property~3 follows from the equality
\beqa
& &\psi_{\r,n,\pb}(\Ltt_1,\Ltt_2(y)-\Ltt_2)\non\\
& &=(2\pi)^{-1/2} \int d\xb\, e^{-i\pb\xb}  (L_1\ti{\Psi}_{\r,n}(\pb)|[(L_2(y)-L_2),(\FF_{\r,n}\otimes I)(-\xb)]\Om)
\eeqa
and from the dominated convergence theorem. Invariance under translations (property~2) is a straightforward consequence of the formula
\beqa
\al_{\x}(L)\ti{\Psi}_{\r,n}(\pb)=e^{-i(\pb,\pb)\x }U_{\pi_{\r}}(x) L\ti{\Psi}_{\r,n}(\pb),\quad \x\in\real^2. \label{translations-formula}
\eeqa
Making use of the above relation and the spectrum condition, it is easy to see that the distribution
\beqa
\real^2\ni q\to (2\pi)^{-1}\int d^2x\, e^{-iqx}\psi_{\r,n,\pb}(\Ltt_1,\al_{x}(\Ltt_2))
\eeqa
is supported in $V_{+}-(\pb,\pb)$. 

Now let us show that any function $\pb\to\psi_{\r,n,\pb}(\,\cdot\,,\,\cdot\,)$, $n\in\Sub_{\r}$,
is non-zero on a non-empty open set $\De_{\r,n}$: Since $(\FF_{\r,n}\otimes I)\Om\in \hil_{\pir}$ is 
different from zero, $U_{\pir}$ does not have non-zero invariant vectors and  
$\pi_{\r}(\Btt)$ acts irreducibly on $\hil_{\pir}$, Lemma~\ref{constant1} ensures the existence of $\Ltt\in\mcLtt$
s.t. $L(\FF_{\r,n}\otimes I)\Om\neq 0$.
Consequently, $\real\ni \xb\to LU_{\pi_{\r}}(\xb)^*(\FF_{\r,n}\otimes I)\Om$ is a non-zero  function 
and so is  the norm of its
Fourier transform $\real_+\ni \pb\to \psi_{\r,n,\pb}(\Ltt,\Ltt)$. Since the functions 
$\real_+\ni \pb\to \psi_{\r,n,\pb}(\Ltt_1,\Ltt_2)$ are
continuous for all $\Ltt_1,\Ltt_2\in\mcLtt$, as we have shown above, the sets
\beq
\De_{\r,n}:=\bigcup_{\Ltt_1,\Ltt_2\in\mcLtt}\{\,\pb\in\real_+\,|\, \psi_{\r,n,\pb}(\Ltt_1,\Ltt_2)\neq 0\,\}
\eeq
are open and non-empty.

Let us now fix some $n\in\Sub_{\r}$, $\pb\in\De_{\r,n}$ and consider the GNS representation $\pi$ induced by $\psi_{\r,n,\pb}$, acting on the  Hilbert space $\hil_{\pi}:=(\mcLtt/\{\, \Ltt\in \mcLtt \,|\, \psi_{\r,n,\pb}(\Ltt,\Ltt)=0 \,  \} )^{\te{cpl}}$.
The equivalence class of $\Ltt\in\mcLtt$ is denoted by $|\Ltt\ran$ and the scalar product is given by 
$\lan \Ltt_1|\Ltt_2\ran=\psi_{\r,n,\pb}(\Ltt_1,\Ltt_2)$. This GNS representation has the form
\beqa
\pi(\Att)|\Ltt\ran&=&|\Att\Ltt\ran, \quad\,\,\,\,\,\, \Ltt\in\mcLtt,\, \Att\in \Btt,\\
U_{\pi}(\x)|\Ltt\ran&=&|\al_{\x}(\Ltt)\ran, \quad \Ltt\in\mcLtt,\,\x\in\real^2,
\eeqa
where $U_{\pi}$ is the standard representation of translations. We will show that $(\pi(\Btt),U_{\pi})$  is unitarily 
equivalent to $(\pi_{\r}(\Btt),U_{\pi_{\r}})$. To this end,
we introduce the map $W_{\r}:\hil_{\pi}\to\hil_{\pir}=\hilk_{1,\odd}\otimes\hilk_{2,\ev}$ 
given by
\beq
W_{\r}|\Ltt\ran=L\ti\Psi_{\r,n}(\pb),\quad \Ltt\in\mcLtt.
\eeq
This map is clearly an isometry. Since $\pi_{\r}$ acts irreducibly on $\hil_{\pir}$, we obtain
that $W_{\r}$ has a dense range and hence it is a unitary operator.  
 From the  relation
\beq
W_{\r}\pi(\Att)|\Ltt\ran=\pi_{\r}(\Att)L\ti\Psi_{\r,n}(\pb)=\pi_{\r}(\Att)W_{\r}|\Ltt\ran,\quad \Ltt\in\mcLtt,\,\, \Att\in\Btt
\label{unitary-equivalence}
\eeq
we conclude that $\pi$ and $\pi_{\r}$ are unitarily equivalent.
Next,  we obtain for any $\Ltt\in\mcLtt$ and $\x\in\real^2$
\beqa
U_{\pi_{\r}}(\x)W_{\r}|\Ltt\ran=e^{i(\pb,\pb)\x} \al_{\x}(L)\ti\Psi_{\r,n}(\pb)=e^{i(\pb, \pb)\x}W_{\r}U_{\pi}(\x)|\Ltt\ran,
\label{intertwining-ev}
\eeqa
where in the first step we made use of relation~(\ref{translations-formula}).
We recall that the spectrum of $U_{\pi_{\r}}$ coincides with $V_+$ and note that $\pi(\Btt)$ acts irreducibly on $\hil_{\pi}$, by  
relation~(\ref{unitary-equivalence}) and Proposition~\ref{asymptotic-representations} (c). Thus, in view of equality~(\ref{intertwining-ev}),
\beq
U^{\can}_{\pi}(\x)=e^{i(\pb, \pb)\x}U_{\pi}(\x),\quad \x\in\real^2 \label{canonical-vs-standard-ev}
\eeq
is the canonical representation of translations in the GNS representation of $\psi_{\r,n,\pb}$. 
Relation~(\ref{canonical-vs-standard-ev}) shows that $q_{\r,n,\pb}=(\pb, \pb)$.

The analysis of the second term on the r.h.s. of (\ref{starting-point}) proceeds similarly: For any $m\in\Sub_{\l}$ and $\Ltt\in\mcLtt$
one introduces vectors
\beq
L\ti{\Psi}_{\l,m}(\pb):=(2\pi)^{-1/2} \int d\xb\, e^{i\pb\xb} LU(\xb)^*(I\otimes \FF_{\l,m})\Om
\eeq
and functionals $\psi_{\l,m,\pb}(\Ltt_1,\Ltt_2)=(L_1\ti{\Psi}_{\l,m}(\pb)|L_2\ti{\Psi}_{\l,m}(\pb))$. By an analogous
reasoning as above one shows that for $\pb$ in some non-empty, open set $\De_{\l,m}\subset\real_+$ these functionals are particle
weights with characteristic energy-momentum vectors $q_{\l,m,\pb}=(\pb, -\pb)$. Their GNS representations are unitarily
equivalent to $\pi_{\l}$. \qed

\section{Conclusions and outlook}\label{conclusions}

In this work we carried out a systematic study of particle aspects of two-dimensional
conformal field theories both in vacuum representations and in charged representations. In the former case
we established a complete particle interpretation in terms of Wigner particles (or `waves' in the terminology of \cite{Bu75}).     
In the latter case we proved the existence of infraparticles and verified  superselection of their
direction of motion in a large class of examples. We conclude that conformal field theories provide a valuable testing ground for
fundamental concepts of scattering theory. 

An important question which remained outside of the
scope of the present work is the problem of asymptotic completeness in the case of infraparticles. We
remark that the theory of particle weights offers natural formulations of this property \cite{Bu87,Bu94} which  
can be adapted to the case of massless, two-dimensional theories. We conjecture that any charged representation of
a chiral conformal field theory has a complete particle interpretation in terms of infraparticles.

A more technical circle of problems concerns the decomposition of particle weights and their representations stated in 
formulas~(\ref{decomposition-representation}), (\ref{decomposition-first}).
We recall that the general procedure of \cite{Po04.1, Po04.2} is not canonical: Firstly, it involves a choice of a
maximal abelian subalgebra, acting on the representation space of the original weight. 
Secondly, it relies on a selection of countable subsets of all the objects involved. 
In view of these ambiguities it is not yet possible to associate
a unique family of (infra-)particle types with any given quantum field theory. We feel that a satisfactory
solution of these problems requires a systematic study of examples. A useful
criterion for their classification is the type of representations induced by particle weights. Thus in the
present paper we focused on representations of type I (with atomic center) which have a simple decomposition
theory. Already in this elementary case we found a physically interesting phenomenon: superselection of direction
of motion. It is a natural direction of further research to look for theories, whose asymptotic functionals induce
representations which are not of type I with atomic center. We conjecture that such models exist and some of them  
describe infraparticles with superselected momentum, similar to the electron in QED.

\vspace{0.5cm}

\noindent\bf Acknowledgements. \rm
The authors would like to thank Prof.$\!$ D.$\!$ Buchholz and Prof.$\!$ R.$\!$ Longo for interesting discussions. 

\vspace{0.5cm}

\appendix

\section{Proof of Theorem~\ref{convergence-of-asymptotic-functionals}}\label{Appendix-A}
\setcounter{equation}{0}

\bel\label{constant1} (a) Let $(\B,U)$ be a local net of $C^*$-algebras on $\real^2$ in
the sense of Definition~\ref{two-dim-net}, acting irreducibly on a Hilbert space $\hil$
and let $U=U^{\can}$. Let $\Psi\in\hil$ be s.t.
\beq
A(f)\Psi:=\int d^2\x\, \al_{\x}(A) f(\x)\Psi=0 \label{AfPsi}
\eeq
for all local operators $A\in\B$ and all $f\in S(\real^2)$ s.t. $\supp\,\tf$ is compact and  $\supp\,\tf\cap V_+=\emptyset$.
Then $\Psi$ is invariant under the action of $U$. (Here $\tf(p):=(2\pi)^{-1}\int d^2x\,e^{ipx}f(x)$ ). \\
(b) Let $(\A,\U)$  be a local net of von Neumann algebras on $\real$ in the sense
of Definition~\ref{circle-theory}, acting irreducibly on a Hilbert space $\K$. Let
$\Psi\in\K$ be s.t.
\beq
A(f)\Psi:=\int d\t\, \alb_{\t}(A)f(\t)\Psi=0
\eeq 
for all local operators $A\in\A$ and all $f\in S(\real)$ s.t. $\supp\,\tf$ is compact and $\supp\,\tf\cap \real_+=\emptyset$. 
Then $\Psi$ is invariant under the action of $\U$. (Here $\tf(\om):=(2\pi)^{-\h}\int ds\,e^{i\om s}f(s)$ ).
\eel
\proof The argument exploits some ideas from the proof of Proposition~2.1 of \cite{BF82}.
As for part (a), suppose that $\Psi$ is not invariant under the action of $U$.  
Since the map $B(\hil)\ni A\to A(f)$ is $\si$-weakly continuous (cf. Lemma~5.3 of \cite{Po04.1})  and $\B$ acts irreducibly
on $\hil$, condition~(\ref{AfPsi}) implies that
\beq
P(\De_1)A(f)P(\De_2)\Psi=0,
\eeq
where $P(\,\cdot\,)$ is the spectral measure of $U$ and $\De_1,\De_2\subset\real^2$ are open bounded sets.
Since the spectrum of $U$ has Lorentz invariant lower boundary and $\Psi$ is not invariant under translations,
we can choose $\De_1,\De_2$ s.t. $P(\De_1)\neq 0$, $P(\De_2)\Psi\neq 0$  and the closure of $(\De_1-\De_2)$
does not intersect with $V_+$. Choosing $f\in S(\real^2)$ s.t.  $\supp\,\tf\cap V_+=\emptyset$ and $\tf(p)=1$ for
$p$ in the closure of $(\De_1-\De_2)$, we obtain that
\beq
P(\De_1)AP(\De_2)\Psi=0
\eeq
for any $A\in\B$. Exploiting irreducibility again, we obtain $P(\De_1)=0$, which is a contradiction.
The proof of part (b) is analogous. \qed
\bel\label{bases} Let $\K_{\pm}\subset\hil_{\pm}$ be closed subspaces, invariant under the action of $U$.  Let $\{e_{+,m}\}_{m\in\II}$ be a complete orthonormal basis in $(P_E\K_+)$ and let $\{e_{-,n}\}_{n\in\JJ}$ be a complete orthonormal basis in $(P_E\K_-)$ for some $E\geq 0$. 
Then any $\Psi\in P_E(\K_+\pout\K_-)$ can be expressed as
\beq
\Psi=\sum_{m,n}c_{m,n}e_{+,m}\pout e_{-,n}, \label{expansion}
\eeq  
where $\sum_{m,n}|c_{m,n}|^2<\infty$.  
\eel
\proof First, we define a strongly continuous unitary representation of translations
\beq
 U_0(x)(\Psi_+\otimes \Psi_-)=(U(x)\Psi_+)\otimes (U(x)\Psi_-),\quad \Psi_{\pm}\in\K_{\pm} 
 \eeq
on $\K_+\otimes \K_-$. Then we obtain from Proposition~\ref{scattering-second-lemma} 
\beq
\Om^{\tout}U_0(x)=U(x)\Om^{\tout}. \label{wave-intertwining}
\eeq
For $\Psi'=(\Om^{\tout})^{-1}\Psi$  the above relation gives $P_{0,E}\Psi'=\Psi'$,
where $P_{0,E}$ is the spectral projection of $U_0$ corresponding to the set $\{\, (\om,\pb)\in\real^2\,|\, \om\leq E\,\}$.  
By the functional calculus, we get $P_{0,E}=P_{0,E}(P_E\otimes P_E)$. Hence
\beq
\Psi'=(P_E\otimes P_E)\Psi'=\sum_{m,n}c_{m,n}e_{+,m}\otimes e_{-,n}.
\eeq
By applying $\Om^{\tout}$, we obtain relation~(\ref{expansion}). \qed
\bel\label{double-commutator-vanishing} Let $\Psi'\in \hil$ be a vector of bounded energy, let $F_1, F_2\in \B$ be almost local and of compact energy-momentum transfer, and let $L=\sum_{k=1}^n A_k B_k$, where  $A_k,B_k\in\B$ are almost local and $B_k$ are, in addition, energy decreasing. Then 
\beq
\lim_{T\to\infty}(\Psi'|[[Q_T,\Phip(F_1)],\Phim(F_2)]\Om)=0, \label{double-commutator}
\eeq
where $Q_T=\int dt\, h_T(t)\int d\xb\, (L^*L)(t,\xb)$. (The above sequence is well defined by Theorem~\ref{integrability}).
\eel
\proof First, we note that by Proposition~\ref{scattering-first-lemma} and Theorem~\ref{integrability}
\beq
\lim_{T\to\infty}(\Psi'|[[Q_T,\Phip(F_1)],\Phim(F_2)]\Om)=\lim_{T\to\infty}(\Psi'|[[Q_T,F_{1,+}(h_T)],F_{2,-}(h_T)]\Om),
\eeq
if the  limit on the r.h.s. exists. We introduce the auxiliary operators:
\beq
Q_{\pm, T}:=\int dt\, h_T(t)\int_{\real_{\pm}} d\xb\, (L^*L)(t,\xb).
\eeq
As we show below, they satisfy
\beq
\lim_{T\to\infty}\|P_E[Q_{\pm, T},F_{\mp}(h_T)]P_E\|=0 \label{commutator-vanishing}
\eeq
for any $E\geq 0$ and any $F\in \B$ which is almost local and of compact energy-momentum transfer. Making use of 
this relation and the fact that $Q_T=Q_{+,T}+Q_{-,T}$, the proof is completed with the help of the Jacobi identity
and Proposition~\ref{scattering-first-lemma}~(c). 

Let us now verify~(\ref{commutator-vanishing}). As the two cases are analogous, we focus on one of them and estimate
the corresponding expression as follows.
\beq
\|P_E[Q_{-, T},F_{+}(h_T)]P_E\|\leq\int dt dt_1\, h_T(t)h_T(t_1)\int_{\real_-}d\xb\|[L^*L(t,\xb), F(t_1,t_1)]\|. \label{commutators}  
\eeq 
Since $L^*L$ and $F$ are almost local, we can find sequences $C_r, F_{r}\in\B(\mco_r)$, 
s.t. for any $n\in \nat$ there exist $C_n, C'_n$ s.t.
\beqa
\|L^*L-C_r\|\leq \fr{C_n}{r^n},\quad \|F-F_{r}\|\leq \fr{C'_n}{r^n}. \label{almost-local-approx}
\eeqa
We choose $r=(1+\fr{1}{4}|\xb|)^{\eps}+T^{\eps}$, where $0<\eps<1$ appeared in the definition of $h_T$. We write
\beqa
[(L^*L)(t,\xb), F(t_1,t_1)]&=&[(L^*L-C_r)(t,\xb), F(t_1,t_1)]\non\\
&+&[C_r(t,\xb), (F-F_{r})(t_1,t_1)]\non\\
&+&[C_r(t,\xb), F_{r}(t_1,t_1)]. \label{many-commutators}
\eeqa
By estimates (\ref{almost-local-approx}), the first two terms on the r.h.s. above give contributions to (\ref{commutators}) which tend to zero in the limit $T\to\infty$. The contribution of the last term can be estimated as follows, exploiting locality,
\beqa
& &\int dt dt_1\, h_T(t)h_T(t_1)\int_{\real_-}d\xb \|[C_r(t,\xb), F_{r}(t_1,t_1)]\|\non\\
& &\ph{444444444444}\leq c\int dt dt_1\, h_T(t)h_T(t_1)\int_{\real_-}d\xb\chi(|\xb-t_1|\leq |t-t_1|+2r),\label{locality-bound}
\eeqa
where $\chi$ is the characteristic function of the respective set and
$c$ is a constant independent of $T$. Let us now derive some inequalities which hold on the support of the integrand on the r.h.s. of 
(\ref{locality-bound}). First, we note that $t,t_1\in \supp\,h_T$, if and only if
$t,t_1\in T^{\eps}\supp\,h+T$, in particular $|t-t_1|\leq c_1T^{\eps}$ for some $c_1\geq 0$. Exploiting this fact, 
the inequality $|\xb-t_1|\leq |t-t_1|+2r$ and the relation
$r= (1+\fr{1}{4}|\xb|)^{\eps}+T^{\eps}$, we find such $c_2\geq 0$ that  $|\xb|\leq c_2T$ and $r\leq c_2T^{\eps}$, 
in particular the r.h.s. of (\ref{locality-bound}) is finite for any $T\geq 1$. Making use of the inequalities  $r\leq c_2T^{\eps}$ and $|\xb-t_1|\leq |t-t_1|+2r$, and of the fact that $t,t_1\in\supp\,h_T$,
 we obtain that $|\xb-T|\leq c_3T^{\eps}$ for some $c_3\geq 0$ which implies that $\xb>0$ for sufficiently large $T$. As the region of  integration in the $\xb$ variable is restricted to $\real_-$, we conclude that the r.h.s. of (\ref{locality-bound}) is zero for such $T$. \qed

\noindent\bf Proof of Theorem~\ref{convergence-of-asymptotic-functionals}: \rm Let $Q_T=\int dt\, h_T(t)\int d\xb\, (L^*L)(t,\xb)$, where 
$L=\sum_{k=1}^n A_k B_k$ is an element of the left ideal $\mcL$, $A_k$ are almost local and $B_k\in\mcL_0$.  
Moreover, we choose  $\Psi=\Phip(F_1)\Phim(F_2)\Om$ and $\Psi'=\Phip(F_1')\Phim(F_2')\Om$, where
$F_{1/2}, F_{1/2}'\in\B$  are almost local and have compact energy-momentum transfer.
Since $\Psi$ and $\Psi'$ are vectors of bounded energy, we can write
\beqa
(\Psi'|Q_T \Phip(F_1)\Phim(F_2)\Om)&=&(\Psi'|[[Q_T,\Phip(F_1)],\Phim(F_2)]\Om)\non\\
&+&(\Psi'|\Phim(F_2)Q_T\Phip(F_1)\Om)\non\\
&+&(\Psi'| \Phip(F_1)  Q_T\Phim(F_2)\Om).\label{identity}  
\eeqa
The term with the double commutator above vanishes as $T\to\infty$
due to  Lemma~\ref{double-commutator-vanishing}. The second term on the r.h.s. of relation~(\ref{identity}) 
is treated as follows: 
\beqa
& &\lim_{T\to\infty}(\Psi'|\Phim(F_2)Q_T\Phip(F_1)\Om)\non\\
& &\ph{44444}=\lim_{T\to\infty}(\Psi'|\Phim(F_2)\int h_T(t)e^{iHt}\int d\xb\, (L^*L)(\xb)e^{-i\Pb t}\Phip(F_1)\Om)\non\\
& &\ph{444444}=\lim_{T\to\infty}(\Psi'|\Phim(F_2)\int h_T(t)e^{i(H-\Pb)t}\int d\xb\, (L^*L)(\xb)\Phip(F_1)\Om)\non\\
& &\ph{4444444444444444444444}=(\Psi'|\Phim(F_2)P_{+}\int d\xb\, (L^*L)(\xb)\Phip(F_1)\Om), 
\eeqa
where in the first step we made use of the fact that $\Phip(F_1)\Om=P_+\Phip(F_1)\Om$, in the second step
we exploited the invariance of $\int d\xb\, (L^*L)(\xb)$ under translations in space and in the last step we
made use of the mean ergodic theorem as in the proof of Lemma~1 of \cite{Bu75}. Next, we obtain
\beqa
& &(\Phip(F_1')\Phim(F_2')\Om|\Phim(F_2)P_{+}\int d\xb\, (L^*L)(\xb)\Phip(F_1)\Om)\non\\
& &\ph{44}=(\Phim(F_2)^*\Phim(F_2')\Om|\Phip(F_1')^*P_{+}\int d\xb\, (L^*L)(\xb)\Phip(F_1)\Om)\non\\
& &\ph{44444}=(\Om|\Phim(F_2')^*\Phim(F_2)\Om)\,(\Om|\Phip(F_1')^*\int d\xb\, (L^*L)(\xb)\Phip(F_1)\Om),
\eeqa
where we made use of the facts that $[\Phip(F_1),\Phim(F_2)]=0$ and that $\hil_+/\CC\Om$ is orthogonal to $\hil_-/\CC\Om$
(as in the proof of Lemma~4 (a) of \cite{Bu75}). The last term on the r.h.s. of (\ref{identity}) is treated analogously.

We note that any $\Psi_{\pm}\in P_E\hil_{\pm}$ can be approximated by a sequence of vectors of the form $P_{\pm}F_n\Om$, where $F_n\in\B$ are
quasilocal and have energy-momentum transfers in some fixed compact set. Hence, any $\Psi=\Psi_{+}\pout\Psi_-$ 
has bounded energy. By the above considerations and  Theorem~\ref{integrability}, 
we obtain for any $\Psi=\Psi_{+}\pout\Psi_-$, $\Psi'=\Psi_{+}'\pout\Psi_-'$, $\Psi_{\pm},\Psi_{\pm}'\in P_E\hil_{\pm}$,
\beqa
\lim_{T\to\infty}(\Psi'|Q_T\Psi)&=&(\Psi_+'|\Psi_+)\int d\xb\, (\Psi_-'|(L^*L)(\xb)\Psi_-)\non\\
&+&(\Psi_-'|\Psi_-) \int d\xb\, (\Psi_+'|(L^*L)(\xb)\Psi_+). \label{preliminary-convergence}
\eeqa
Now in view of Lemma~\ref{bases}, any $\Psi\in P_E\hil^\tout$ has the form 
\beq
\Psi=\sum_{m,n}c_{m,n}e_{+,m}\pout e_{-,n}, \label{general-psi}
\eeq
where $\{e_{\pm,m}\}_{m=0}^{\infty}$ are  orthonormal systems in $\{P_E\hil_\pm\}$, which we choose so that $e_{\pm,0}=\Om$. 
Defining
\beqa
 \Psi_{+,n}=\sum_{m}c_{m,n}e_{+,m},\quad
 \Psi_{-,n}=\sum_{m}c_{n,m}e_{-,m},  \label{constituting-vectors}
\eeqa
we obtain $\rho_{\pm,\Psi}(\,\cdot\,)=\sum_n\,(\Psi_{\pm,n}|\,\cdot \, \Psi_{\pm,n})$. 
Relation~(\ref{preliminary-convergence}) gives  
\beq
\lim_{T\to\infty}(\Psi|Q_T\Psi)=\int d\xb\,(\rho_{+,\Psi}+\rho_{-,\Psi}) \big( (L^*L)(\xb)\big). \label{preliminary-convergence-two}
\eeq
Exploiting the Cauchy-Schwarz inequality and the following bounds, valid for $L=AB$, $A\in\mfa$, $B\in\mcL_0$,  
\beqa
|(\Psi|Q_T\Psi)| &\leq& \|P_{E}\int d\xb\, (B^*B)(\xb) P_{E}\|\, \|A^*A\|, \\
\int d\xb\,(\rho_{+,\Psi}+\rho_{-,\Psi}) \big( (L^*L)(\xb)\big)&\leq& 2\|P_E\int d\xb\, (B^*B)(\xb) P_E\|\,\|A^*A\|,\,\,\,\, 
\eeqa
one extends (\ref{preliminary-convergence-two}) to any $L\in\mcL$. Now formula~(\ref{main-technical}) follows
by a polarization argument. 

Let us now show that $\psi_{\Psi}^{\tout}=0$  only if  $\Psi\in \CC\Om$. By the above considerations
we obtain, for any  $B\in\mcL_0$, 
\beq
\psi_{\Psi}^{\tout}(B,B)=\sum_n\int d\xb\,\big\{ (\Psi_{-,n}|(B^*B)(\xb)\Psi_{-,n})
+(\Psi_{+,n}|(B^*B)(\xb)\Psi_{+,n}) \big\}.
\eeq
If $\psi_{\Psi}^{\tout}=0$, then  $B\Psi_{\pm,n}=0$ for each $n$ and any such $B$. 
Thus, by Lemma~\ref{constant1} (a),  $\Psi_{\pm,n}$ are proportional to $\Om$. Using
definitions~(\ref{general-psi}), 
(\ref{constituting-vectors}) and the convention $e_{\pm,0}=\Om$,  it is easily seen that $\Psi$ is proportional to $\Om$. \qed


\section{Proofs of Lemmas~\ref{odd-lemma} and \ref{even-lemma}    } \label{Appendix-B}
\setcounter{equation}{0}

\bf Proof of Lemma~\ref{odd-lemma}: \rm  As for the main part of the lemma, it suffices to show that the spectrum of $\U_{\odd}$ coincides
with $\real_+$. It follows from the assumption $\Ad W \neq \id$ and the Reeh-Schlieder property of the net $(\A,\U)$ 
that $\A_{\odd}(\I)\neq \{0\}$ and $\K_{\odd}=[\A_{\odd}(\I)\Omz]$ for any open, bounded subset $\I\subset\real$.  
Let $P(\,\cdot\,)$ be the spectral measure of $\U$ and suppose that $P(\De)\K_{\odd}=\{0\}$ for some 
open subset $\De\subset\real_+$. We fix a non-zero $A\in \A_{\odd}(\I)$. Then, for any $B\in  \A(\I)$ the distribution
\beq
(\Omz|[B,\wt{A}(\om)]\Omz)=\fr{1}{\sqrt{2\pi}}\int dt\, e^{-i\om t}(\Omz|[B,\be_t(A)]\Omz)
\eeq  
is supported outside of $\De \cup -\De$. Since, by locality, this distribution is a holomorphic function, it must be zero
for all $\om\in\real$. Thus for any $f\in S(\real)$ s.t. $\tf$ is supported in the interior of $\real_+$ we obtain
$(\Omz|BA(f)\Omz)=(\Omz|B\tf(T)A\Omz)=0$. Here $T\geq 0$ is the generator of $\U$, $A(f)=\int dt\, \be_t(A)f(t)$ and
we made use of the fact that  $A(f)^*\Omz=0$, due to the support property of $\tf$ and the spectrum condition. 
Approximating the characteristic function of the interior of 
$\real_+$ with such $\tf$ and making use of the fact that $(\Omz|A\Omz)=0$, we conclude that $A\Omz=0$ and hence, by
the Reeh-Schlieder property  $A=0$, which contradicts our assumption. Consequently, $P(\De)\K_{\odd}\neq\{0\}$ for
any open subset $\De$ of $\real_+$, which means that the spectrum of $\U_{\odd}$ coincides with $\real_+$.

This fact can also be proven as follows: The representation of translations $V$ can be extended to a
representation of the $ax+b$ group thanks to the Borchers theorem \cite{Fl98}. There is only one non-trivial, irreducible
representation of this group  which has positive energy \cite{Lo08} and its spectrum of translations is $\RR_+$. 
Since $\K_{\odd}$ does not contain non-trivial invariant vectors of $V$, the spectrum of $V|_{\K_{\odd}}$ coincides
with $\real_+$.

Let us now proceed to part (a) of the lemma. To show the irreducibility of $\pi_{\odd}$, it suffices to check that any vector $\Psi\in\hilk_{\odd}$ is cyclic under the action of $\pi_{\odd}(\A_{\ev})$.
By contradiction, we suppose that there is $\Psi'\in\hilk_{\odd}$ s.t. $(\Psi'|A\Psi)=0$ for any $A\in\A_{\ev}$. But this
implies that $(\Psi'|B\Psi)=0$ for any $B\in\A$, which contradicts the irreducibility of the action of $\A$ on $\hilk$.  
Next we verify the faithfulness of $\pi_{\odd}$ restricted to a local algebra. Let $A \in \A_\ev(\I)$ be a positive local element which is zero upon restriction to $\hilk_\odd$. For any local odd element $B \in \A_\odd(\J)$ and
for sufficiently large $s$ we obtain
\beq
0 = (\Om|\be_{s}(B^*)A \be_s(B)\Om)=(\Om|\be_s(B^*B)A\Om) \to (\Om|B^*B\Om)\cdot(\Om|A\Om),
\eeq
where in the last step we took the limit $s \to \infty$.
By the Reeh-Schlieder property it follows that $A=0$.  This implies that
$\pi_\odd$ is faithful on $\A_{\ev}(\I)$ by Proposition~2.3.3~(3) of \cite{BR}.
Now the faithfulness of $\pi_{\odd}$ on the quasilocal algebra $\A_{\ev}$ 
follows from Proposition~2.3.3~(2) of \cite{BR},
which says that $\pi_{\odd}$ is faithful, if and only if $\|\pi_{\odd}(A)\|=\|A\|$ for any $A \in \A_{\ev}$.
Local normality of $\pi_{\odd}$ is obvious, since $\pi_{\odd}$ acts by the restriction to a subspace. Indeed,
making use of  Lemma 2.4.19 from \cite{BR} and of the fact that $\pi_\odd$ preserves the norm,
it is easy to check that $\te{l.u.b.} \pi_{\odd}(A_{\al})= \pi_{\odd}(\te{l.u.b.}\, A_{\al})$, where l.u.b denotes the least upper
bound and  $\{A_{\al}\}_{\al\in \II}$ is a uniformly bounded increasing net of positive operators from some $\A_{\ev}(\I)$.

Part (b) of the lemma follows from the uniqueness of the invariant vector of $\U$. \qed


\noindent\bf Proof of Lemma~\ref{even-lemma}: \rm  We know from Lemma~\ref{odd-lemma} that $\A_{\ev}\neq \CC I$, since it can be irreducibly represented
on the infinite dimensional Hilbert space $\K_{\odd}$. Consequently, we can find a non-zero $A\in\A_{\ev}(\I)$,
for some open, bounded $\I$, s.t. $(\Omz|A\Omz)=0$. Proceeding identically as in the proof of the main part of Lemma~\ref{odd-lemma},
we conclude that the spectrum of $\U_{\ev}$ coincides with $\real_+$.  Part (b) follows trivially from the fact that the net $(\A, V)$
is in a vacuum representation. Irreducibility in part (a) follows from part (b). The remaining part of the statement is 
proven analogously as the corresponding part of Lemma~\ref{odd-lemma}. \qed

\end{document}